\documentclass[11pt]{article}
\pdfoutput=1

\usepackage{mheck}

\usepackage{amsmath,amssymb,epsfig,amsfonts}
\usepackage{graphicx,subfigure}
\usepackage{cite}
\usepackage{verbatim} 
\usepackage{float}
\restylefloat{table}
\usepackage{pdflscape}
\usepackage{tikz}
\usepackage{array}
\usepackage{youngtab}
\usepackage{multirow}
\usepackage{color}
\usepackage{ulem}

\makeatletter

\definecolor{darkbrown}{rgb}{0.4, 0.26, 0.13}
\definecolor{greenb}{rgb}{0.0, 0.7, 0.2}

\newcommand{\be}{\begin{equation}}
\newcommand{\ee}{\end{equation}}
\newcommand{\ba}{\begin{aligned}}
\newcommand{\ea}{\end{aligned}}

\newcommand{\bea}{\begin{eqnarray}}
\newcommand{\eea}{\end{eqnarray}}

\def\ch{\hbox{ch}}
\def\points{x}

\def\unit{{1\kern-.65ex {\rm l}}}
\def\1{{1\kern-.65ex {\rm l}}}

\def\bbP{\mathbb{P}}

\def\CN{{\cal N}}
\def\CO{{\cal O}}

\def\bbP{{\mathbb{P}}}

\def\bbS{{\mathbb{S}}}

\newcount\hour \newcount\minute
\hour=\time \divide \hour by 60
\minute=\time
\def\now{%
\ifnum \hour<13
  \ifnum \hour=0 \advance \hour by 12 \number\hour:\else \number\hour:\fi%
     \ifnum \minute<10 0\fi%
     \number\minute%
\ A.M.%
\else \advance \hour by -12 \number\hour:%
  \ifnum \minute<10 0\fi%
  \number\minute%
  \ P.M.%
\fi%
}

\makeatother

\newcommand{\het}{X} 

\newcommand{\hets}{X_{19,19}} 

\newcommand{\hbun}{{\mathcal E}} 

\newcommand{\hbe}{B_\het} 

\newcommand{\Fthg}{Y} 

\newcommand{\Fth}{Y_{\rm DGW}} 
\newcommand{\Fbe}{B_{\rm DGW}} 
\newcommand{\Fpr}{\pi}

\newcommand{\Mth}{J} 

\institution{OXFORD}{\ Mathematical Institute, Oxford University, Woodstock Road, Oxford OX2 6GG, UK }
\institution{SCGP}{\ Simons Center for Geometry and Physics, SUNY, Stony Brook, NY 11794, USA}
\institution{NEU}{\ Department of Physics, Northeastern University, Boston, MA 02115, USA}
\institution{UPSALLA}{\ Department of Physics and Astronomy, Uppsala University, SE-751 20 Uppsala, Sweden}
\institution{UCSB}{\ Department of Mathematics, University of California, Santa Barbara, CA 93106, USA}

\title{\Huge  Infinitely Many M2-instanton Corrections to  M-theory on $G_2$-manifolds}

\authors{
Andreas P. Braun \worksat{\OXFORD}, 
Michele Del Zotto \worksat{\SCGP},
James Halverson \worksat{\NEU}, \\
\bigskip
Magdalena Larfors \worksat{\UPSALLA}, 
David R. Morrison \worksat{\UCSB}, 
Sakura Sch\"afer-Nameki \worksat{\OXFORD} \\
\bigskip
}

\abstract{\noindent 
We consider the non-perturbative superpotential for a class of four-dimensional $\mathcal N=1$ vacua obtained from M-theory on seven-manifolds with holonomy $G_2$. The class of $G_2$-holonomy manifolds we consider are so-called twisted connected sum (TCS) constructions, which have the topology of a K3-fibration over $S^3$. We show that the non-perturbative superpotential of M-theory on a class of TCS geometries receives infinitely many inequivalent M2-instanton contributions  from infinitely many three-spheres, which we conjecture  are supersymmetric (and thus associative) cycles. 
The rationale for our construction is provided by the duality chain of \cite{Braun:2017uku}, which relates M-theory on TCS $G_2$-manifolds to $E_8\times E_8$ heterotic backgrounds on the Schoen Calabi-Yau threefold, as well as to F-theory on a K3-fibered Calabi-Yau fourfold.  The latter are known to have an infinite number of instanton corrections to the superpotential and it is these contributions that we trace through the duality chain back to the $G_2$-compactification.
}

\date{}
\begin{document}

\maketitle

\tableofcontents

\section{Introduction}

Four-dimensional superstring vacua that preserve minimal supersymmetry are among the most interesting both theoretically and phenomenologically. The heterotic superstrings of type $E_8 \times E_8$ or $\text{Spin}(32)/\mathbb{Z}_2$ compactified on a Calabi-Yau (CY) threefold $\het$ together with an appropriate choice of stable holomorphic gauge bundle, $\hbun$, give a well-known method to generate examples of this sort. Other well-known instances of backgrounds of this kind are F-theory models associated to\footnote{These also have an interpretation as the type IIB string compactified on the base of the elliptic fibration, with the fibers specifying the variable axio-dilaton field, see, e.g., \cite{whatF}.}
elliptically fibered CY fourfolds $\Fthg$ together with four-form flux, or by M-theory on $G_2$ holonomy seven-manifolds $\Mth$. 
The least well-understood of these is M-theory on $G_2$ holonomy manifolds, largely due to the difficulty in constructing and studying compact geometries of this type. 
Recently, however, 
a large class of compact, smooth $G_2$-manifolds were obtained as twisted connected sums (TCS) \cite{MR2024648, Corti:2012kd, MR3109862}.
Remarkably, a subclass of {M-theory compactifications on TCS $G_2$-manifolds} are connected by dualities to heterotic and F-theory compactifications \cite{Braun:2017uku}. Essential for {these dualities} is that each TCS geometry
comes equipped with a K3-fibration, which in turn allows a fiber-wise application of M-theory/heterotic duality, and subsequently heterotic/F-theory duality. 
The main aim of this paper is to gain insight into the physics of the {M-theory compactification} by exploiting this duality chain in order to identify infinitely many
non-perturbative superpotential contributions that are known to exist in the F-theory compactification \cite{Donagi:1996yf}.


M-theory compactifications on $G_2$ holonomy manifolds have the rather unique feature of being largely geometric. This has to be contrasted to the other known examples of 4d $\CN =1$ vacua, in which the compactification geometry needs to be supplemented with additional data. For instance, in the case of an F-theory background this includes the choice of a four-form flux as well as the presence of space-time filling 
D3-branes, required to cancel the tadpole that arises {in the case of} non-vanishing Euler characteristic {of the total space} \cite{Sethi:1996es,Gukov:1999ya}. The presence of these additional structures often complicates 
identifying the origin of various physical effects in the 4d effective theory, which explains {one of the advantages} of working with $G_2$-compactifications in M-theory.  
{However this simplification comes with the price that the geometry} of {$G_2$ holonomy}  manifolds is much more complicated than that of complex Calabi-Yau varieties, which are amenable to algebro-geometric tools. 
Not surprisingly, our guide {to understanding these manifolds} is precisely the string duality we alluded to above.

A large class of compact $G_2$ holonomy manifolds have {recently} been constructed by {Corti, Haskins, Nordstr\"om, and Pacini} \cite{Corti:2012kd, MR3109862}, building upon {earlier} work by Kovalev \cite{MR2024648}. Some aspects of the physics of these so-called twisted-connected sum (TCS) $G_2$-manifolds have been explored in the context of M-theory \cite{Halverson:2014tya,Halverson:2015vta,Guio:2017zfn, Braun:2017uku} and superstring \cite{Braun:2017ryx,Braun:2017csz} compactifications. {A} key feature of these backgrounds is that TCS $G_2$-manifolds are topologically K3-fibrations over a three-sphere. This structure is suggestive of fiberwise M-theory/heterotic duality, and indeed it was shown that a subclass of TCS $G_2$-manifolds are dual to heterotic compactifications on the Schoen Calabi-Yau threefold $\hets$ \cite{Braun:2017uku}. Since these heterotic models are among the best studied 4d $\CN=1$ backgrounds, this duality gives a natural framework to overcome the difficulties arising from the lack of {algebro-geometric tools} on the $G_2$ side. The TCS $G_2$-manifolds considered in \cite{Braun:2017uku} are the ideal framework to explore the non-perturbative physics of M-theory compactifications to four-dimensions.

Consider a heterotic $E_8 \times E_8$ compactification. {Despite being well-studied,} {very little is known} about which pairs $(\het,\hbun)$ give rise to consistent ${\mathcal N}=1$ heterotic backgrounds: while it is possible to find pairs  $(\het,\hbun)$  that solve the classical equations of motion at every order in $\alpha^\prime$, these can be destabilized non-perturbatively by world-sheet instantons \cite{Dine:1986zy,Dine:1987bq}. Often while being individually non-trivial, the sum of the contributions from all world-sheet instantons vanishes \cite{Distler:1986wm,Distler:1987ee,Silverstein:1995re,Beasley:2003fx}. Nevertheless certain world-sheet instantons give contributions that cannot cancel against each other and therefore give rise to a non-perturbative superpotential that, for appropriate bundle data, never vanishes (see e.g.\ \cite{Buchbinder:2016rmw,Buchbinder:2017azb} for two recent works about this phenomenon). One of the best known examples of this sort is provided by the so-called $E_8$-superpotential of Donagi--Grassi--Witten (DGW) originally computed in F-theory \cite{Donagi:1996yf} and later mapped to a {dual} heterotic compactification on the Schoen Calabi-Yau \cite{Curio:1997rn}. In that context one has a superpotential that receives infinitely many possible contributions of which only a fraction at a time can vanish, depending on the bundle data and on the presence and location of space-time filling wrapped NS5-branes. The goal of this paper is to trace through the duality chain, and identify these DGW superpotential contributions in terms of M2-branes wrapped on  three-cycles in the $G_2$ holonomy manifold.

For M-theory compactifications on $G_2$-manifolds the question of non-perturbative corrections is 
{equally poorly understood}.
While classically these backgrounds are stable, non-perturbatively generated superpotentials could destabilize these vacua. In this context, known contributions to the superpotentials are generated by Euclidean M2-branes (EM2) wrapped on associative three-cycles of $J$ that are rational homology three-spheres \cite{Harvey:1999as}. It is well-known that associative cycles have an obstructed deformation theory, and therefore are not stable under variations of the $G_2$-structure of a given $G_2$-manifold \cite{Mclean96}, the latter {corresponding to moving in} the moduli space of M-theory. This feature of the associative cycles is the key for reproducing correctly the corresponding behavior of the superpotentials that we have mentioned briefly above. While in F-theory or in heterotic string theory the vanishing of such terms is associated to non-geometric properties, e.g.\ to a Ganor zero in F-theory \cite{Ganor:1996pe}, in M-theory this is due to the moduli-dependent existence of the corresponding associative three-cycles. Our task is then to identify via the duality map an infinite number of  three-cycles that give rise to the analog of the DGW superpotential in the M-theory compactification.  Based on the duality, we conjecture that the three-cycles we find have associative representatives.

\begin{figure}
\begin{center}
\includegraphics[width=4.4cm]{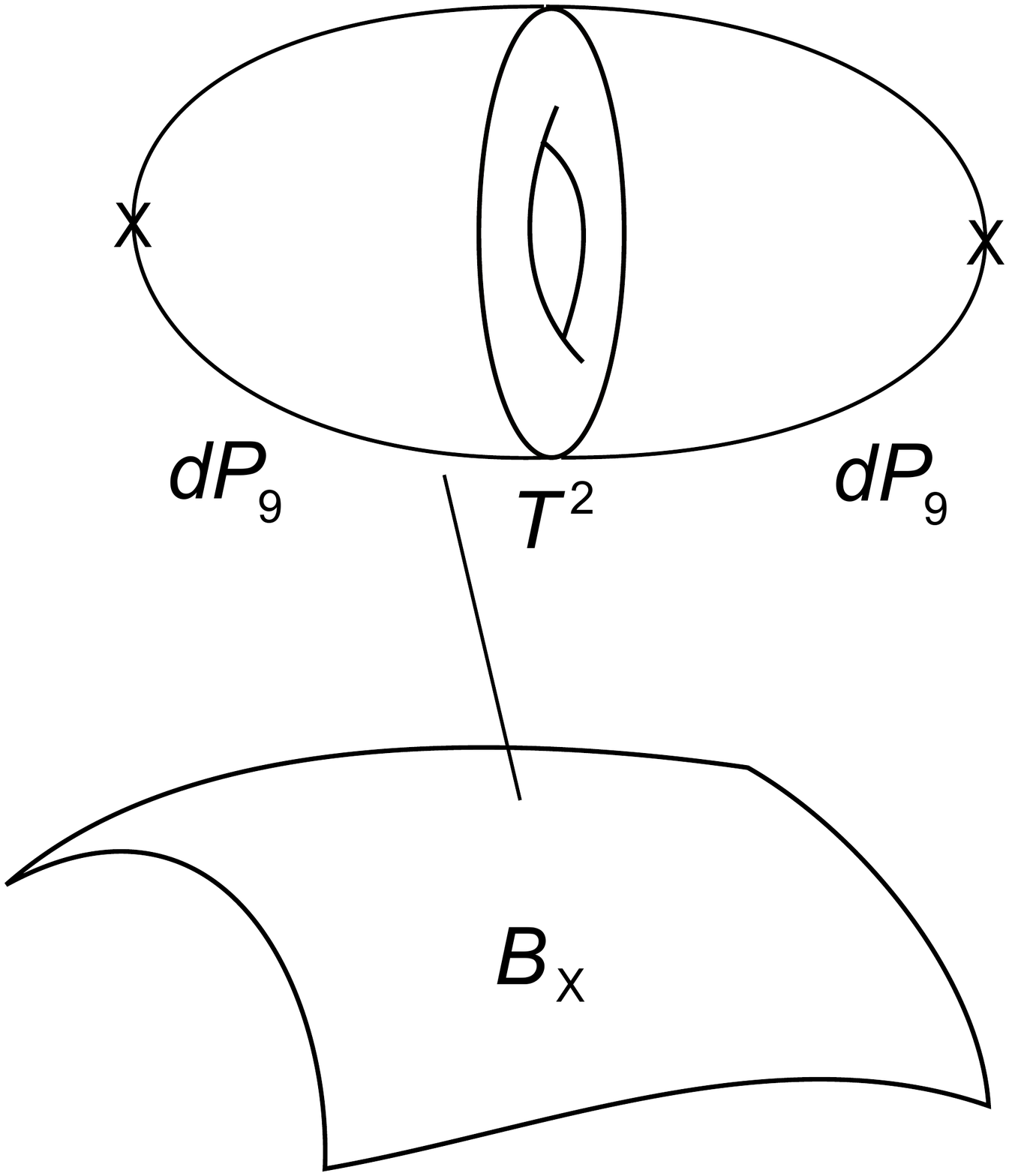}\qquad \qquad\qquad \qquad 
\includegraphics[width=4.4cm]{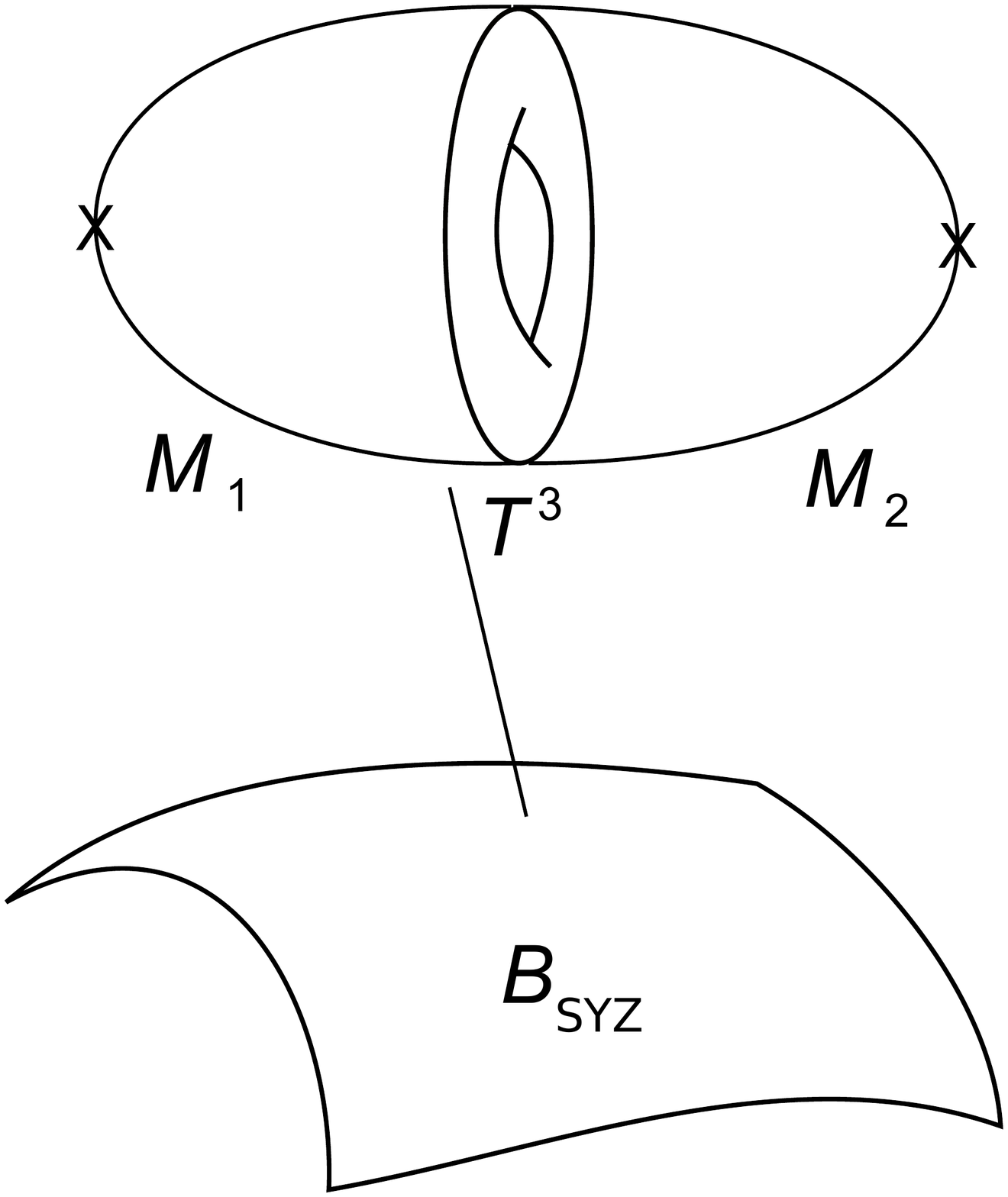}
\caption{LHS: F-theory/Heterotic duality and stable degeneration limit. The F-theory fourfold is realized as a K3-fibration over the same base {as the elliptic fibration $\het \to \hbe$ of the heterotic Calabi-Yau threefold}. The {heterotic bundle data}  are summarized by the moduli of the two $dP_9$ surfaces that are glued along a $T^2$, which is identified with the elliptic fiber of $\het$. RHS:
M-theory/Heterotic duality and analogue of the stable degeneration limit. The M-theory $G_2$-background is realized as a K3-fibration over the same base {as} the heterotic SYZ-fibration. The {heterotic bundle data}  are summarized by the moduli of the two {``half-K3''} four-manifolds, $M_1$ and $M_2$, that {are glued} along a $T^3$, which is identified with the SYZ-fiber of $X$ \cite{DaveStrings2002}.}\label{Het_M}
\end{center}
\end{figure}

Whenever a heterotic CY threefold $\het$ admits a Strominger-Yau-Zaslow (SYZ) fibration by special Lagrangian three-tori, it is possible to {apply} a fiberwise M-theory/heterotic duality to map {$\het$} to a K3-fibered $G_2$-manifold $\Mth$  \cite{Strominger:1996it, Gukov:2002jv}. This {suggests} \cite{DaveStrings2002}  that an analogue of the stable degeneration limit for the F-theory fourfolds should exist also for the $G_2$-manifolds that are dual to heterotic (see figure  \ref{Het_M}). This is precisely the case for the Schoen Calabi-Yau,  and it is possible to represent it in terms of a connected-sum type construction, which is naturally dual to the TCS-construction of $G_2$-manifolds \cite{Braun:2017uku}. 
It is then possible to match the world-sheet instantons on the heterotic side to M2-brane instantons on the M-theory side and identify dual three-cycles in the TCS $G_2$-manifold, which we conjecture to have supersymmetric (i.e., associative) representatives.
In this process, we find that in the SYZ-description of the Schoen Calabi-Yau, the holomorphic cycles corresponding to the world-sheet instantons look like thimbles that are glued together into two-spheres by a matching condition on the $S^3$ base of the SYZ-fibration. Under the duality the {circle-}fiber of the thimble is replaced by an $S^2$ and each thimble is thus mapped to a half-$S^3$. The matching condition responsible for gluing the thimbles into $S^2$s is dualized to a matching condition that glues the half-$S^3$s into $S^3$s. However, we find that the matching condition on the M-theory side is more refined than that on the Schoen, and it is supplemented with extra geometric data that is keeping track e.g.\ of the positions of the heterotic space-time filling wrapped NS5-branes. In this way we identify the  three-cycles that are needed to reconstruct the DGW superpotentials on the M-theory side. We conjecture that these are new calibrated three-cycles in this class of TCS $G_2$-manifolds, which give rise to {infinitely many contributions} to the superpotential. That these are associative cycles is inferred indirectly via the duality: the curves in heterotic and surfaces in F-theory  are supersymmetric, whereby the expectation is that these newly identified three-cycles in the $G_2$-manifold should also have calibrated representatives with are rational homology three-spheres.

Our paper is organized as follows. Section \ref{sec:FHG} is devoted to the starting point for our analysis, which is the chain of dualities from F-theory via heterotic string theory to M-theory on a $G_2$ holonomy manifold, focusing on the examples investigated in \cite{Braun:2017uku}. In section \ref{sec:instantonreview}, after a brief review of the DGW superpotential in F-theory and its dual heterotic version on the Schoen Calabi-Yau,  the string junction picture for the heterotic world-sheet instantons in the SYZ-description is discussed, which is crucial for the duality map to  M-theory. Section \ref{sec:G2Insts} is the core of the paper where we lift the string junction picture from heterotic to M-theory exploiting the TCS construction of the backgrounds investigated in \cite{Braun:2017uku}. In particular, in section \ref{sec:g2lift} we present our conjectures regarding the existence of infinitely many associative three-cycles on TCS $G_2$-manifolds. In section \ref{sec:outlook} we summarize our results and discuss directions for future studies.  Several technical details are discussed in the Appendices.


\section{Twisted Connected Sum $G_2$-manifolds and Dualities}
\label{sec:FHG}

This section gives a brief summary of the duality chain of \cite{Braun:2017uku}. 
The starting point  is the construction of twisted connected sum (TCS) $G_2$ holonomy manifolds \cite{MR2024648, Corti:2012kd, MR3109862, Braun:2017uku}, which naturally come equipped with a K3-fibration. 
The duality between M-theory on K3 and heterotic on $T^3$ can be applied fiberwise resulting in a duality 
between M-theory on a TCS $G_2$-manifold $J$, and heterotic $E_8 \times E_8$ string theory on an SYZ-fibered Calabi-Yau threefold $X$. The additional structure required for the duality to be explored in detail is that the K3-fibers in the TCS construction are {themselves} elliptically fibered. In this case, the Calabi-Yau threefold in the heterotic dual is the Schoen threefold \cite{Schoen} {or split bi-cubic} \cite{Candelas:1987kf}. 
{Furthermore, this heterotic compactification can be obtained by stable degeneration from the F-theory model associated to a K3-fibered Calabi-Yau fourfold.}
We first briefly summarize the TCS-construction and then provide further details on the duality chain.

\subsection{TCS-construction of $G_2$-manifolds}
\label{sec:FHGsub}

For future reference we introduce some more notation for the TCS construction. A TCS $G_2$-manifold is constructed from 
two building blocks $Z_\pm$, which are algebraic threefolds $Z_\pm$ with a K3-fibration over $\mathbb{P}^1$. The K3 fibers can be thought of as elements in a lattice polarized family of K3 surfaces, and we denote a generic K3 fiber, i.e., a generic element of this lattice polarized family, by $S_\pm$. Crucially, the building blocks have a non-vanishing first Chern class, which is equal to the class {of} a K3 fiber 
\be
c_1 (Z_\pm) = [S_\pm] \,,
\ee
and satisfy $h^{i,0}(Z_\pm) = 0$ for $i\neq 0$. Fixing a generic fiber $S_\pm^0$, this implies that $X_\pm = Z_\pm \setminus S_\pm^0$ are asymptotically cylindrical Calabi-Yau threefolds, i.e., there is a Ricci-flat metric of holonomy $SU(3)$ on $X_\pm$ and outside of a compact subset $X_\pm$ are isomorphic to a product\footnote{  We denote circles by $\mathbb{S}^1$ to avoid confusion with the surfaces $S$.} $\mathbb{R}_+ \times \mathbb{S}^1_{b \pm} \times S_\pm^0$. 

\begin{figure}
\begin{center}
\includegraphics[width=9cm]{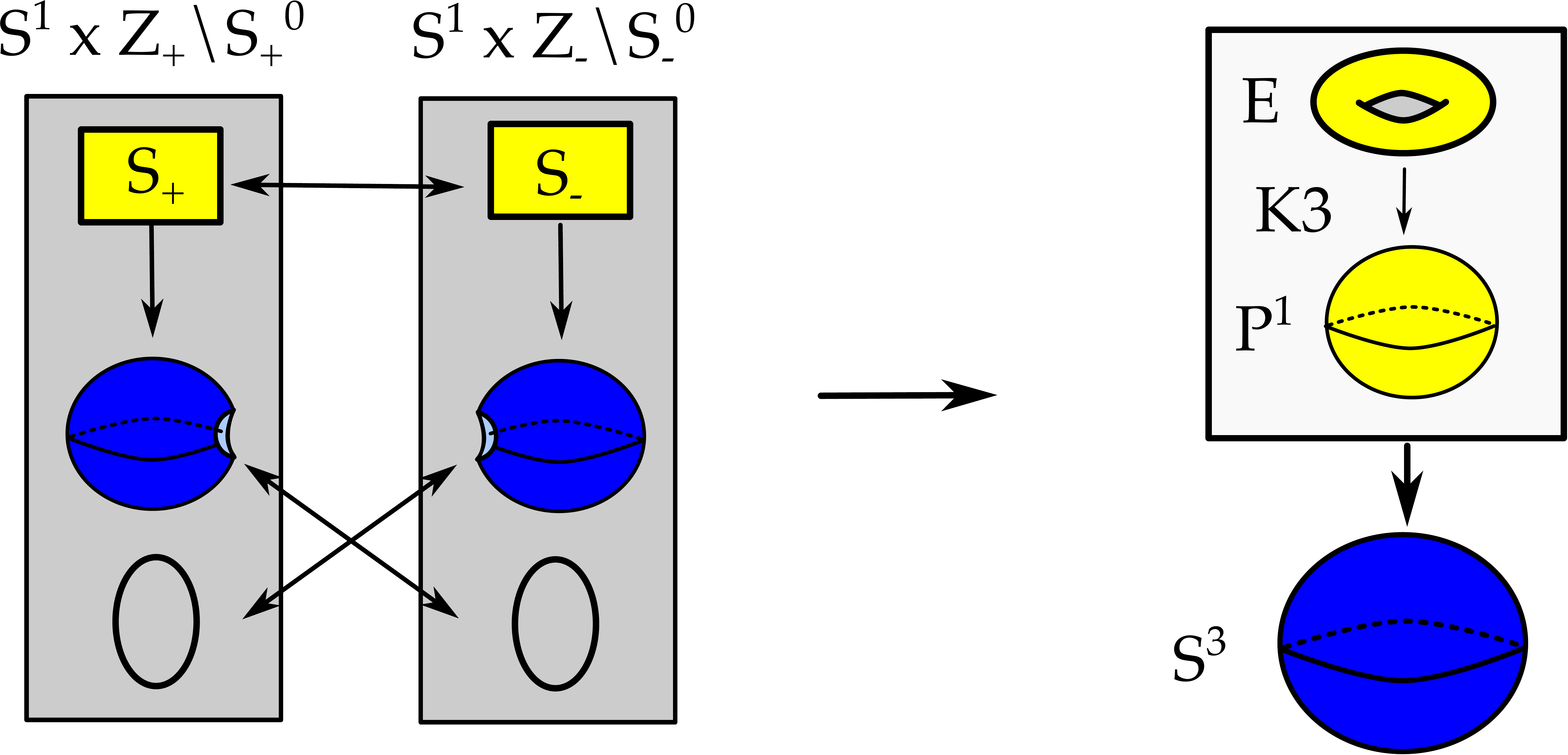}
\caption{The Twisted Connected Sum construction for the $G_2$ holonomy manifold $J$. The left hand side shows the building blocks $Z_\pm$, which are K3-fibered over an open $\mathbb{P}^1$s.  The gluing {involves} a hyper-K\"ahler rotation and exchange of $\mathbb{S}^1_{e\pm}$ with $\mathbb{S}^1_{b\mp}$ along the cylindrical central part. The global structure of the TCS manifold is that of a K3-fibration over $S_3$, as shown on the right hand side. {For the duality chain to be applicable, we require the K3-surfaces in each building block to be elliptically fibered.} 
\label{fig:TCS}}
\end{center}
\end{figure}

A TCS $G_2$-manifold $J$ is then found by gluing $X_\pm \times \mathbb{S}^1_{e\pm}$ along their cylindrical regions by identifying
$\mathbb{S}^1_{e\pm}$ with $\mathbb{S}^1_{b\mp}$ and mapping $S_+^0$ to $S_-^0$ by a hyper-K\"ahler rotation $\phi$. In particular, $\phi$ is chosen such that it maps $\omega_\pm$ to Re$\,\Omega^{(2,0)}_{\mp}$, as well as Im$\,\Omega^{(2,0)}_+ \leftrightarrow -$Im$\,\Omega^{(2,0)}$.  Following \cite{Halverson:2014tya}, we refer to $\phi$ as a Donaldson matching. A sketch is shown in figure \ref{fig:TCS}.

The second cohomologies of the K3-fibers can be decomposed as 
\be
H^2 (S_\pm, \mathbb{Z}) \cong \Lambda = U_1 \oplus U_2 \oplus U_3 \oplus (-E_8) \oplus (-E_8) \,,
\ee
where we have labeled the three summands of the hyperbolic lattice\footnote{ The hyperbolic lattice is the unique even two-dimensional lattice of signature $(1,1)$. There exists a basis of generators with inner {product matrix} $\begin{pmatrix}0& 1 \\ 1& 0\end{pmatrix}$\,.} $U$ by an index $i=1, \cdots, 3$. 
There is a natural restriction map 
\be
\rho_\pm :\qquad H^2 (Z_\pm, \mathbb{Z}) \rightarrow H^2 (S_\pm, \mathbb{Z})\,,
\ee
which allows us to define the lattices 
\be
N_\pm = \hbox{im} (\rho_\pm)\,,\qquad 
K(Z_\pm)= 
\ker(\rho_\pm)/[S_\pm] \,.
\ee
The polarizing lattices of the K3 fibers $S_\pm$ contain (and in many cases are equal to) the lattices $N_\pm$, which must
be primitively embedded in $H^2 (K3, \mathbb{Z})$. The orthogonal complement of $N_\pm$ in $H^2 (S_\pm, \mathbb{Z})$ is 
\be
T_{\pm} = N_{\pm}^\perp \subset H^2 (S_\pm, \mathbb{Z})  \,.
\ee
The Donaldson matching $\phi$ implies an isometry $H^2 (S_+, \mathbb{Z}) \cong H^2 (S_-, \mathbb{Z})$, which in turn defines a common embedding 
\begin{equation}
N_\pm \ \hookrightarrow\  \Lambda \, . 
\end{equation}
Conversely, given such embeddings of $N_\pm$, we may find an associated Donaldson matching if {there is} a compatible choice of the forms 
$\omega_\pm$ and $\Omega^{(2,0)}_{\pm}$ for fibers $S_\pm^0$ in the moduli space of the algebraic threefolds $Z_\pm$.

With this information on the matching, the integral cohomology of $J$ can be determined using the Mayer-Vietoris exact sequence as
\be\label{CohomologiesTCS}\ba
H^1(J,\mathbb{Z}) & =   0 \cr 
H^2(J,\mathbb{Z}) & =  (N_+ \cap N_-) \ \oplus \ K(Z_+) \ \oplus\  K(Z_-) \cr 
H^3(J,\mathbb{Z}) & = 
\mathbb{Z}[S] \ \oplus\  \Gamma^{3,19} /(N_+ + N_-) \ \oplus\  (N_- \cap T_+)\  \oplus\  (N_+ \cap T_-)\cr 
&\quad  \oplus H^3(Z_+)\ \oplus \ H^3(Z_-) \ \oplus\  K(Z_+) \ \oplus \ K(Z_-) \cr 
H^4(J,\mathbb{Z}) & = 
H^4(S) \oplus 
(T_+ \cap T_-)\  \oplus \ \Gamma^{3,19} /(N_- + T_+) \ \oplus\  \Gamma^{3,19} /(N_+ + T_-) \cr 
& \quad  \oplus  H^3(Z_+)\oplus H^3(Z_-) \ \oplus \ K(Z_+)^* \oplus K(Z_-)^* \cr 
H^5(J,\mathbb{Z}) & = \Gamma^{3,19} /(T_+ + T_-) \ \oplus K(Z_+) \oplus K(Z_-) \,.
\ea\ee
We refer the reader for a more in depth discussion of these geometries to \cite{MR2024648, Corti:2012kd, MR3109862, Braun:2017uku}. 

We can now describe the geometry that will be central to the present paper, {which} was initially discussed in \cite{Braun:2017uku}. For this smooth TCS $G_2$-manifold, the lattices $N_\pm$ and $T_\pm$ for the generic K3-fibers of the building blocks are chosen as follows
\be
\ba
N_+ & = U_2 \cr 
N_- & = U_3 \oplus (-E_8) \oplus (-E_8) 
\ea 
\qquad 
\ba
T_+ &= U_1 \oplus U_3 \oplus (-E_8) \oplus (-E_8) \cr 
T_- &= U_1 \oplus U_2 \,.
\ea
\ee
This implies that the K3-fibers $S_+$ and $S_-$ are elliptically fibered and that the elliptic fibration of $S_+$ is given by a generic (smooth) Weierstrass model over $\mathbb{P}^1$, whereas the elliptic fibration of $S_-$ has two $II^*$ fibers. We have anticipated a Donaldson matching by a labeling of the various summands of $U$ lattices, which implies in particular that 
\be
N_+ \cap N_- = \{\vec{0}\} \quad N_+\cap T_- = U_2, \quad N_-\cap T_+ = U_3 \oplus (-E_8) \oplus (-E_8) \quad \text{and}\quad T_+ \cap T_- = U_1.
\ee
The explicit algebraic realization of the building blocks $Z_\pm$ is discussed in some more detail in section \ref{sec:G2Insts}. The relevant topological data are 
\begin{equation}
\ba
h^{1,1}(Z_+) &= 3\cr 
h^{2,1}(Z_+) &= 112 \cr 
|K_+| & = 0 
\ea
\qquad 
\ba
h^{1,1}(Z_-) & = 31\cr 
h^{2,1}(Z_-)& = 20 \cr 
|K_-| & = 12 
\end{aligned}\,.
\end{equation}
It is now straightforward to apply \eqref{CohomologiesTCS} to find the Betti numbers of the associated smooth TCS $G_2$-manifold $J$ as 
\begin{equation}\label{eq:bettiJ}
 b_2(J) = 12 \hspace{2cm} b_3(J) = 299 \, .
\end{equation}
In conclusion, the spectrum of M-theory compactified on {this} TCS $G_2$-manifold $J$ consists of 12 vectors and 299 chiral multiplets in 4d.

\subsection{Duality Chain: M-theory/Heterotic/F-theory}
\label{sec:MHF}

The K3-fibration that TCS $G_2$-manifolds automatically come equipped with is rather suggestive in terms of applications to M-theory compactifications and string dualities. The duality of M-theory on K3 and heterotic on $T^3$ is based on the observation that the moduli spaces of both compactifications are given by
\be
\Gamma \backslash  SO(3,19)/(SO(3)\times SO(19) ) \times \mathbb{R}^+ \,,
\ee
which serves as both the moduli space of Einstein metrics on K3 and the Narain moduli space of heterotic strings on $T^3$.  The $\mathbb{R}^+$ represents the volume modulus for the K3 surface and is also identified with the heterotic string coupling. In \cite{Braun:2017uku} it was proposed to apply M-theory/heterotic duality fiberwise to TCS $G_2$-manifolds, resulting in heterotic string theory on Calabi-Yau threefolds, which are $T^3$-fibered. The application of this fiberwise duality is straightforward if the K3 fibers $S_\pm$ furthermore carry elliptic fibrations. {In a nutshell  the duality chain implies an equivalence between the following 4d $\mathcal{N}=1$ string vacua.}
The M-theory compactification on the TCS $G_2$-manifold $J$ is dual to a heterotic $E_8\times E_8$ string compactified on the Schoen (or ``{split} bi-cubic'') Calabi-Yau threefold $\hets$ with vector bundles whose data is specified in terms of the TCS geometry. Generically, these bundles completely break the $E_8\times E_8$ gauge symmetry. On the other hand, the Schoen Calabi-Yau threefold has an elliptic fibration with base $dP_9$, so that heterotic string theory on the Schoen Calabi-Yau is dual to F-theory associated to a Calabi-Yau fourfold given as a K3-fibration over $dP_9$. This is precisely the Calabi-Yau threefold studied by Donagi--Grassi--Witten (DGW) \cite{Donagi:1996yf}. The idea in this paper is to follow the non-perturbative superpotential contributions computed in \cite{Donagi:1996yf} back through this duality chain and identify these contributions in the M-theory on TCS $G_2$s.

\begin{figure}
\begin{center}
\includegraphics[width=16cm]{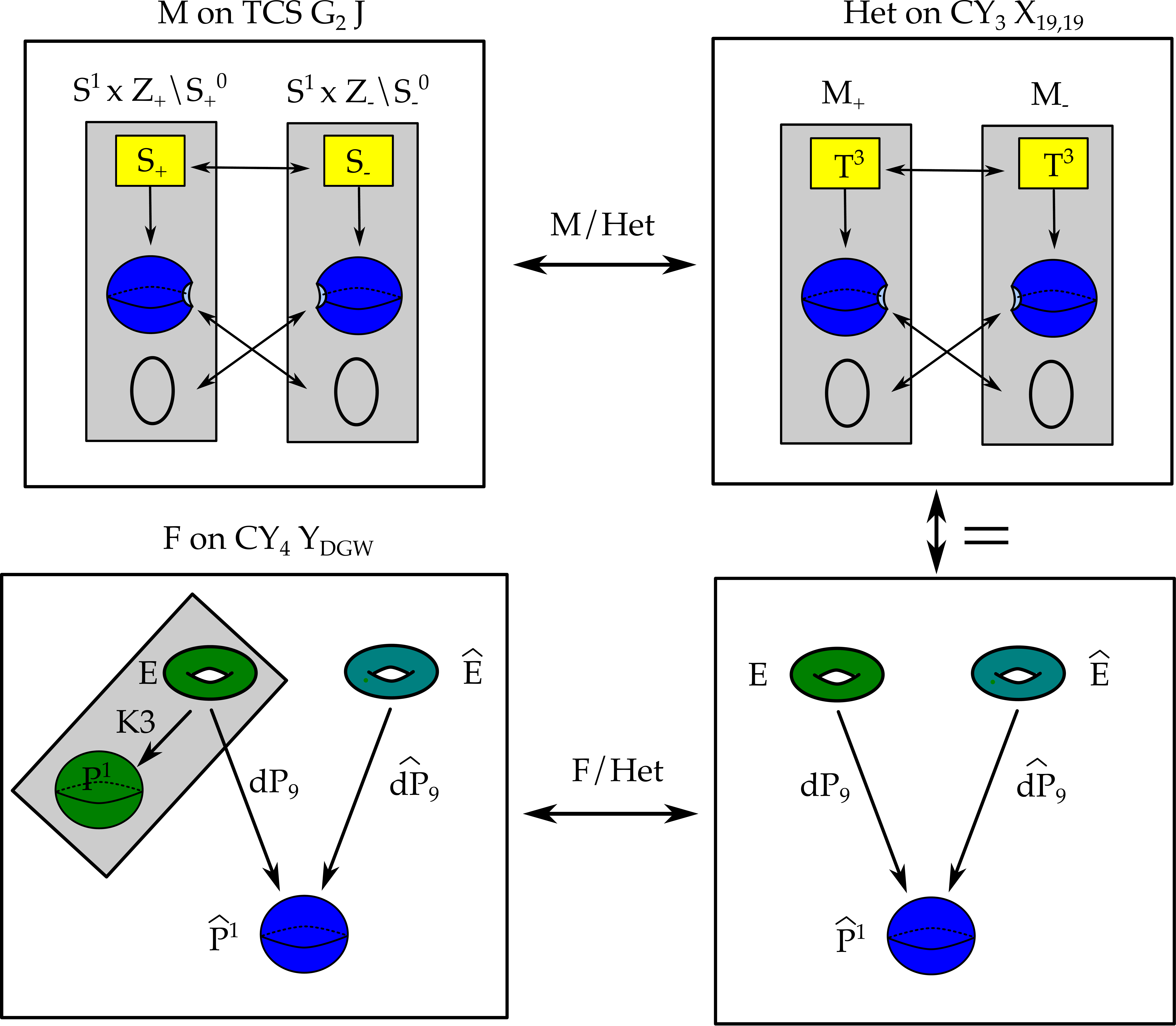}
\caption{Depiction of the duality chain (from upper left in clock-wise direction). The TCS $G_2$-manifold $J$ is K3 fibered over $S^3$ and can be decomposed into a TCS such that both building blocks are fibered by K3 surfaces $S_\pm$, which are themselves elliptically fibered. The building blocks are $Z_{\pm} \backslash S^0_{\pm}$.
 In the dual heterotic string theory the K3 surfaces $S_\pm$ are replaced by three-tori $T^3$, which results in the Schoen $\hets$ Calabi-Yau threefold written as an SYZ-fibration (top right). An alternative description of the Schoen Calabi-Yau is in terms of a double-elliptic fibration {over $\widehat{\mathbb{P}^1}$ (bottom right)}, and application of F-theory/heterotic duality maps this to the elliptic K3-fibered Calabi-Yau fourfold $Y_{\rm DGW}$ studied by Donagi--Grassi--Witten (bottom left).
\label{fig:DualityChain}}
\end{center}
\end{figure}

{To conclude the discussion about the duality chain, let us provide below some more details on the steps involved, referring to \cite{Braun:2017uku} for a more complete discussion.} We have summarized the relevant geometries  in figure \ref{fig:DualityChain}.

\subsection*{M-theory on the TCS $G_2$-manifold $J$ and Heterotic on  $\hets$} 

It follows from the Betti numbers \eqref{eq:bettiJ} that M-Theory compactified on $J$ gives a 4d $\mathcal{N}=1$ theory with $12$ $U(1)$ vector multiplets and $299$ uncharged chiral multiplets. As reviewed above, the TCS $G_2$-manifold $J$ is constructed from two building blocks $Z_\pm$ with elliptic K3 fibers $S_\pm$. As $Z_\pm$ are algebraic, only the complex structures, i.e., $\Omega^{2,0}_\pm$ vary holomorphically over the base $\mathbb{P}^1$s.  After gluing $X_\pm \times \mathbb{S}^1_{e \pm}= Z_\pm \setminus S^0_\pm \times \mathbb{S}^1_{e \pm}$ to form $J$, these K3-fibrations glue to a {\it non-holomorphic} fibration of K3 surfaces over $S^3$. The various degenerations of the K3 fibers over the base $S^3$ of $J$ translate to the combined data of geometry (in the form of the SYZ-fibration) and bundles on the heterotic side by applying fiberwise duality. First of all, this implies that the dual Calabi-Yau geometry $X$ on the heterotic side enjoys a similar `TCS' decomposition as the $G_2$-manifold we started from \cite{Braun:2017uku}. This means we can cut it into two pieces $M_\pm$, such that $X = M_+\cup M_-$ and the complex threefolds $M_\pm$ are fibered by three-tori $T^3$. However, as both $S_\pm$ are elliptically fibered, only a $T^2 \subset T^3$ varies non-trivially over the base of $M_\pm$ and one identifies 
$M_\pm =  V_\pm \times \mathbb{S}^1_{s\pm} \times \mathbb{S}^1_{e\pm}$, where $V_\pm$ are isomorphic to $dP_9\setminus T^2$ as real manifolds. The $T^3$ fibers of the SYZ fibration on $M_\pm$ are given by a product of the elliptic fibers of $V_\pm$ times $\bbS^1_{s\pm}$.
The gluing between $M_\pm$ is induced by the Donaldson matching, which in turn implies that the geometry {$X$} on the heterotic side is given by the Schoen Calabi-Yau $X=\hets$. This construction shows the structure of $\hets$ from the point of view of its SYZ-fibration. Alternatively, $X_{19,19}$ can be viewed as fibration of a product of elliptic curves $\mathbb{E} \times \widehat{\mathbb{E}} $ over a rational curve $\widehat{\mathbb{P}^1}$. The second Chern class of $\hets$ is  
\begin{equation}
c_2( \hets ) =  12(\mathbb{E} +\widehat{\mathbb{E}})\, .
\end{equation}
To have a consistent heterotic compactification, this class must equal the sum of the second Chern character of the $E_8\times E_8$ vector bundle $\hbun$ together with the classes of NS5-branes. 

This data is encoded in the $G_2$-manifold $J$ as follows. The choice $N_- \supset (-E_8)\oplus (-E_8)$ and $T_+ \supset (-E_8)\oplus (-E_8)$ implies that all of the bundle data are carried by $Z_+$. On the heterotic side, this translates to the $E_8\times E_8$ vector bundle $\hbun= \hbun_1 \oplus \hbun_2$ being chosen such that $ch_2(\hbun_1) = ch_2(\hbun_2) = 6\,\hat{\mathbb{E}} $. Furthermore, there are $12$ degenerations of the K3 fiber $S_-$ on $Z_-$ which correspond to $12$ NS5-branes wrapped on $\mathbb{E}$. Altogether, this permits the computation of the spectrum of massless $\mathcal N=1$ multiplets on the heterotic side. There are $12$ $U(1)$ vectors and $3\cdot 12$ complex scalars associated with the $12$ NS5-branes on $\mathbb{E}$. Furthermore, there are $19+19$ moduli from the geometry and $2\cdot 112$ moduli associated with the bundle $\hbun$. Together with the dilaton, this reproduces the spectrum of the dual M-theory compactification.

\subsection*{Heterotic on $\hets$ and F-theory associated to $Y_{\rm DGW}$}

The Calabi-Yau threefold $\hets$ carries an elliptic fibration\footnote{ In fact, this geometry has infinitely many elliptic fibrations \cite{Anderson:2017aux}.} with fiber $\mathbb{E}$ and we are considering a heterotic background with a bundle $\hbun$, which is flat on $\mathbb{E}$ and completely breaks the gauge group $E_8\times E_8$. The base of the elliptic fibration is a rational elliptic surface $dP_9$. This allows us to immediately write down the dual F-theory geometry $Y_{\rm DGW}$ as a generic elliptic fibration over {$B_{DGW} =\mathbb{P}^1 \times  \widehat{dP_9}$}, which is the fourfold considered by Donagi--Grassi--Witten in \cite{Donagi:1996yf}. The relevant topological data of $Y_{\rm DGW}$ are
\begin{equation}
h^{1,1}(Y_{\rm DGW}) = 12\,, \hspace{.5cm} h^{2,1}(Y_{\rm DGW}) = 112\,, \hspace{.5cm} h^{3,1}(Y_{\rm DGW}) = 140\,, \hspace{.5cm} \chi(Y_{\rm DGW}) = 288\, .
\end{equation}
In the dual F-theory, the $12$ NS5-branes on $\mathbb{E}$ become $12$ space-time filling D3-branes, which precisely matches the D3-brane tadpole constraint $\chi(Y_{\rm DGW})/24 = N_{D3} =12 $. These give rise to $12$ $U(1)$ vectors together with $36$ complex scalars in the low-energy effective action. The geometry then contributes $h^{1,1}(B_{DGW}) +  h^{2,1}(Y_{\rm DGW}) +  h^{3,1}(Y_{\rm DGW}) =  11 + 112 + 140 = 263$ complex scalar moduli. Together this again reproduces the spectrum initially found on the M-theory side. 
In our construction, both the building block $Z_+$ and the elliptic fourfold $Y_{\rm DGW}$ are only determined once the distribution of $ch_{2}(\hbun) = ch_{2}(\hbun_1) +  ch_{2}(\hbun_2) = 12(\hat{\mathbb{E}})$ between the two $E_8$ factors $V_1$ and $V_2$ is fixed. 

The geometries we have discussed, which are such that $Z_+$ is elliptically fibered over $\mathbb{P}^1 \times \mathbb{P}^1 $ and $Y_{\rm DGW}$ is elliptically fibered over the base {$B_{DGW} =\mathbb{P}^1 \times \widehat{dP_9}$}, correspond to the symmetric choice $\ch_2(\hbun_1) =  \ch_2(\hbun_2) = 6(\widehat{\mathbb{E}})$. Other choices $Z_{+,n}$ which are elliptic fibrations over the Hirzebruch surfaces $\mathbb{F}_n$ for $n=0, \cdots, 6$ are possible and give rise to geometrically non-Higgsable gauge groups $D_4, E_6, E_7, E_8, E_8$ for $n=2,3,4,5,6$ throughout the duality chain \cite{Braun:2017uku}. This may be generalized to arbitrary elliptic building blocks $Z_+$ (keeping $Z_-$ fixed), the dual F-theory geometry of which can be directly constructed as an elliptic fibration (with fiber $\widehat{\mathbb{E}}$) over $Z_+$.


\section{Instanton Corrections in F-theory and Heterotic}\label{sec:instantonreview}

Using the duality chain reviewed in the last section, and summarized in figure \ref{fig:DualityChain}, we now aim to identify non-perturbative superpotential contributions to M-theory on the TCS $G_2$-manifold $J$. 
The starting point is the observation in Donagi--Grassi--Witten \cite{Donagi:1996yf} that there is  an infinite sum of contributions to the superpotential in F-theory associated to $Y_{\rm DGW}$, due to D3-instantons. We  {shall start} with a summary of their analysis in section \ref{sec:DGW} and first utilize the duality map to heterotic on the Schoen Calabi-Yau (lower half of figure \ref{fig:DualityChain}) \cite{Curio:1997rn} to identify the world-sheet instanton corrections dual to these D3-branes. 
The goal is to follow the duality chain all the way to M-theory on $J$, and identify the dual M2-brane instanton contributions in section \ref{sec:G2Insts}. However before this can be done, the heterotic world-sheet instantons need to first be identified in terms of the SYZ-fibration of the Schoen (upper right corner of figure \ref{fig:DualityChain}) \cite{Braun:2017uku}, which has a direct dual interpretation in the M-theory on $G_2$ compactification. This is done in section \ref{sec:StringJunk} from a string junction point of view.

\subsection{D3-Instantons in F-theory associated to $Y_{\rm DGW}$}
\label{sec:DGW}

Consider the F-theory model associated to an elliptically fibered Calabi-Yau fourfold $\Fth$, with base $\Fbe$ and projection map $\Fpr: \Fth \rightarrow \Fbe$. In the absence of four-form flux, a necessary condition for a divisor $D$ in $\Fth$ to contribute to the superpotential is that \cite{Witten:1996bn}
\be
\chi (D, \mathcal{O}_D) = 1 \,.
\ee
A sufficient condition is that $h^i(D) = 0$, for $i=1,2,3$. Furthermore the only divisors in an elliptic fibration which can contribute are of vertical type, i.e., pull-backs of divisors $D_B$ from the base $\Fbe$, $D=\Fpr^{-1}(D_B)$. For vertical divisors the Euler characteristic is 
\be
\chi (D,\mathcal{O}_D) = - {1\over 24} D\cdot D \cdot c_2 (\Fth) \,,
\ee
which requires in particular that $D\cdot D <0$. As discussed in \cite{Ganor:1996pe}, the contribution of these instantons has the form
\begin{equation}
G(m) \times \text{exp}\left(-V(D_B) + i \, \int_{D_B} C_4^{\,+} \right).
\end{equation}
The prefactors $G(m)$ depend on all the moduli of the problem and account for extra zero-modes that can kill a given contribution to the superpotential. In particular the terms $G(m)$ are sections of the line bundles $[D]$ dual to divisors $D$ and holomorphic sections of line bundles that have no poles must have a simple zero on a manifold homotopic to $D$  \cite{Ganor:1996pe}. 
Assuming that $D$ is isolated ($h^3(D) =0$), for instance, entails that $G$ is zero everywhere along $D$. This has a simple physical explanation. For elliptic fourfolds with nonzero Euler characteristic and in the absence of fluxes, the D3-brane tadpole implies the presence of spacetime filling D3-branes. Each of these D3-branes have a moduli space that equals the fourfold $\Fth$. Whenever one of these D3-branes hits one of the wrapped Euclidean D3-branes {(ED3)} that give rise to the instanton contributions, extra zero modes are generated, which lift that contribution from the potential.

The instanton contributions for F-theory associated to $Y_{\rm DGW}$ were determined in \cite{Donagi:1996yf}. The geometry, as we summarized in the last section, is a K3-fibered Calabi-Yau fourfold, whose base threefold is 
$\Fbe = \widehat{dP_9} \times \mathbb{P}^1$, where the rational curve is the base of the elliptic K3 surface. 
The vertical divisors are pull-backs of base divisors 
\be\label{DivsF}
D_{\Fbe} =  \sigma \times \mathbb{P}^1 \,,
\ee
where the $\sigma$ are irreducible curves in the del Pezzo surface, given in terms of sections of the fibration 
{$p: \widehat{dP_9} \rightarrow \widehat{\mathbb{P}^1}$} satisfying $\sigma^2= -1$. 

Let us describe these sections explicitly. The rational elliptic surface $\widehat{dP_9}$ is elliptically fibered over $\widehat{\mathbb{P}^1}$ with $12$ reducible fibers, in the notation of section \ref{sec:MHF} 
\be
\widehat{\mathbb{E}}\  \hookrightarrow \ \widehat{dP_9}\  \stackrel{p}{\longrightarrow} \  \widehat{\mathbb{P}^1}\,,
\ee
where we will denote the class of the fiber by $[\widehat{\mathbb{E}}] = \widehat{E}$. 
Sections of this fibrations can be identified with the $E_8$ root lattice by noting that the middle cohomology is
\begin{equation}\label{eq:dp9innerform}
H^2(dP_9,\mathbb{Z}) = \begin{pmatrix}
                        -1 &1 \cr   1 & 0
                       \end{pmatrix}\oplus (-E_8)\, .
\end{equation}
Here, the two-dimensional sub-lattice corresponding to the first summand is generated by the fiber $\widehat{E}$ and a choice of zero-section $\sigma_0$, which obey
$\widehat{E}^2 = 0$, $\widehat{E}\cdot \sigma_0=1$ and $\sigma_0^2 = -1$, the latter following from adjunction and the fact that $c_1(dP_9) = [\widehat{E}]$. 
The second summand is the $E_8$ root lattice $(-E_8)$, which can be constructed using string junctions between the $12$ singular fibers. 
Equivalently, it can be derived as follows: for an elliptic surface $S$ with a section, the middle cohomology always takes the form $H^2(S,\mathbb{Z})= \widehat{U} \oplus W$, where $\widehat{U}$ is generated by fiber and zero-section and $W$ is the frame lattice. In the present case, adjunction together with Poincar\'e duality shows that $W$ is an even self-dual lattice, and the signature theorem determines its signature to be $(0,8)$, so that we can conclude that $W = -E_8$. 

As shown in \cite{Donagi:1996yf}, every curve in $H_2(dP_9,\mathbb{Z})$, which squares to $-1$ and meets the fiber $\widehat{\mathbb{E}}$ in a single point is a section of the elliptic fibration. By exploiting this fact we can immediately see the isomorphism between the group of sections and the $E_8$ lattice. Consider a lattice vector $\gamma$ in $E_8$ such that $\gamma^2 = -2n$. Any such vector satisfies $\gamma\cdot \widehat{E} =0$. The corresponding section can be constructed by
\begin{equation}\label{eq:sigmaandgamma}
 \sigma_\gamma \equiv \gamma + \sigma_0 + n \widehat{E} \, ,
\end{equation}
and it is easy to see that
\be\label{SectionCond}
\sigma_\gamma^2 =-1 \qquad \hbox{and}\qquad 
\sigma_\gamma \cdot \widehat{E}=1 \,.
\ee 
Note that the latter fixes the coefficient of $\sigma_0$ to be $1$ and the above becomes the unique form of any curve with the desired properties. Hence there is a unique section corresponding to each element of $E_8$. As $\sigma_0 \sim 0$ in the (additive) group of sections, i.e., the Mordell-Weil group, we hence find that the isomorphism between the group of sections and the free abelian group $\mathbb{Z}^8$, expressed as the lattice $-E_8$. 

We can now use the above description to recover the infinite contribution to the superpotential in \cite{Donagi:1996yf}. For every section $\sigma_\gamma$ there is an associated divisor {$D_B^{\gamma}$} of  {$\Fbe$}  and the superpotential is computed as
\be\label{eq:superpotential_first_principle}
W =  \sum_{\gamma} G_\gamma \,  \, {\text{exp}\left(2\pi i \int_{D_B^\gamma} i \, J_B \wedge J_B + \, C_4^{\,+} \right)} \ ,
\ee
where $J_B$ is the K\"ahler form of $\Fbe$.
To evaluate the sum, we parameterize the Poincar\'e dual of {$i \, J_B \wedge  J_B +  \, C_4^{\,+}$} in terms of 
\begin{equation}
PD( i \, J_B \wedge J_B +  \, C_4^{\,+})  = \sum_k \omega_k C_k \,,
\end{equation}
where $\omega_k\in {\mathbb{C}}$ and $C_k$ are curves on {$\Fbe$}. The only curves for which \eqref{eq:superpotential_first_principle} is non-zero, come from the $dP_9$ in $\Fbe$, so that $k = 0, \cdots ,9$. It is useful to choose a basis  $C_0 =  (\sigma_0 + F )$, $C_9 = \widehat{E}$ and $C_i = \alpha_i^*$ with  $\alpha_i^*\cdot \alpha_j = \delta_{ij}$, where the $\alpha_j$ are a set of simple roots for the $E_8$ in \eqref{eq:dp9innerform}. Furthermore any $\gamma$ in \eqref{eq:sigmaandgamma} can be expanded in terms of the simple roots in $(-E_8)$ as 
\begin{equation}
\gamma =  \sum_{m \in \mathbb{Z}^8} m_i \alpha_i \,,
\end{equation}
which together with $\gamma^2 = - 2n$ and  an appropriate labeling of the simple $E_8$ roots implies that 
\begin{equation}\label{eq:ndoesnotmatter}
n = -\tfrac12\gamma^2 =    \sum_{i=1}^8 ( m_i^2 ) - (m_1 m_2 + \cdots m_6 m_7 + m_3 m_8) \,.
\end{equation}
The expression \eqref{eq:superpotential_first_principle} can now be evaluated 
\begin{equation}
\begin{aligned}
S &= \sum_{\gamma \in E_8} G_\gamma \exp \left[2\pi i \left(\gamma + \sigma_0 + n \widehat{E} \right) \cdot \left(\tau (\sigma_0 + \widehat{E} ) + \widehat{E} \omega_9 + \sum_{i=1}^8 \omega_i \alpha_i^* \right) \right]\\
 &= \sum_{m \in \mathbb{Z}^8} G_m \exp \left[2\pi i  \left(\omega_9 +n \tau + \sum_{i=1}^8 m_i \omega_i \right)\right] \\
  &=  e^{2\pi i \omega_9} \sum_{m \in \mathbb{Z}^8} G_m \exp\left[2\pi i\left( \sum_{i=1}^8 (m_i \omega_i + m_i^2 \tau) - (m_1 m_2 + \cdots m_6 m_7 + m_3 m_8)  \tau \right) \right] \,.
\end{aligned}
\end{equation}
Setting all the prefactors $G_\gamma =1$ reproduces the $E_8$ theta-function $\Theta_{E_8}(\tau,{\bf \omega})$  found in \cite{Donagi:1996yf} after rescaling the K\"ahler parameters $\omega_i$ and $\tau$ by $2\pi i$. 
Note that the structure of the $E_8$ lattice only enters in a rather indirect way through the map \eqref{eq:sigmaandgamma}. For every choice of basis of $H^2(\Fbe)$ there is a dual basis of curves to be used in the expansion \eqref{eq:superpotential_first_principle}. However, the $E_8$ lattice appearing in \eqref{eq:dp9innerform} is \emph{not} mapped to a sublattice of $H^2(\Fbe,\mathbb{Z})$ by \eqref{eq:sigmaandgamma}, which results in the specific form of the terms proportional to $n$ to ultimately lead to the function $\Theta_{E_8}(\tau,{\bf \omega})$. 

We should pause here and discuss the universality of the prefactors $G_\gamma$. In  \cite{Donagi:1996yf}, it was argued that there exists for every pair of sections an automorphism of $\widehat{dP_9}$, which exchanges them. This lifts to a birational automorphism of  $Y_{\rm DGW}$, however the integral in (\ref{eq:superpotential_first_principle}) is independent of this. Therefore one could expect that the coefficients $G_\gamma$ do not depend on $\gamma$. In appendix \ref{sec:universalprefactorevidence} we provide a discussion of the 3--7 zero modes and necessary conditions for a universal prefactor, which are satisfied in this case. However more importantly, due to the non-vanishing Euler characteristic of $Y_{\rm DGW}$ and absence of fluxes, a consistent F-theory compactifications will require spacetime-filling D3-branes. 
These can give rise to Ganor strings \cite{Ganor:1996pe}  that depend on the positions of such D3-branes, which 
generically break the automorphism above, thus destroying the universality of $G_\gamma$. Irrespective of this, there is an infinite sum contributing to the superpotential, which we now map to the heterotic dual, and subsequently to M2-brane instantons in M-theory on the TCS $G_2$-manifold.

\subsection{Heterotic Duality and Worldsheet Instantons}\label{sec:hetdual}

In this section we turn to the heterotic dual picture and identify the  counterparts to the D3-brane instantons in F-theory. These arise from dual heterotic world-sheet instanton contributions, which for the Schoen Calabi-Yau have already been discussed in  \cite{Curio:1997rn}, albeit again neglecting the potential non-universality of the prefactors. (A more recent discussion of a subset of the instantons can be found in section 4.2.2 of \cite{Anderson:2015yzz}.) As explained in section \ref{sec:MHF} the heterotic dual to F-theory associated to $Y_{\rm DGW}$ is compactified on the Schoen Calabi-Yau threefold $\hets$. For the analysis in this section it is most useful to view the Schoen as a double-elliptic fibration over $\widehat{\mathbb{P}}^1$, or equivalently the fiber product $\hets= dP_9 \times_{\widehat{\mathbb{P}}^1} \widehat{dP_9}$. We shall denote the two rational elliptic surfaces by $S$ and $\widehat{S}$, respectively.

Let us first recap when heterotic world-sheet instantons contribute \cite{Witten:1999eg}. For reasons related to holomorphy \cite{Dine:1986zy,Dine:1987bq} only genus zero curves can contribute to the superpotential. Moreover, we are going to consider contributions from instantons that are isolated and smooth (which should coincide with a genericity assumption). The fact that the instantons are isolated translates into a condition of rigidity for the corresponding curve: for an instanton that contributes to the superpotential, the only allowed bosonic zero modes correspond to translations along ${\mathbb R}^{1,3}$, e.g.\ $(-1,-1)$ curves. 

Each such curve $C$ contributes to the superpotential a summand \cite{Witten:1999eg}
\begin{equation}\label{eq:heterotic_instanton_pref}
{\text{Pf} ({\mathcal D}_F) \over  \sqrt{\text{det}({\mathcal D}_B)}} \, \text{exp}\left( - {A(C) \over 2 \pi \alpha^\prime} + i \int_C B \right),
\end{equation}
where ${\mathcal D}_F$ and ${\mathcal D}_B$ are the kinetic operators for the fermionic and the bosonic degrees of freedom of the instanton and $A(C)$ denotes the volume of $C$ as measure by the K\"ahler form.
 The latter can be translated in differential geometric properties of $(\het,\hbun)$. 
In particular, ${\mathcal D}_F$ coincides with the $\bar \partial$ operator on $\hbun\otimes {\mathcal O}(-1)$. If this operator has a nontrivial kernel the Pfaffian in  \eqref{eq:heterotic_instanton_pref} vanishes and the corresponding curve does not contribute to the superpotential. Therefore, {since the dimension of the kernel may increase
on subloci in moduli space} each contribution depends explicitly, via its prefactor, on the bundle data for the given heterotic model.

Naively, in this context one should have a 2d $(0,2)$ sigma-model description, and hence a vanishing criterion for the non-perturbative superpotential \cite{Silverstein:1995re}. The latter has been translated into the Beasley-Witten residue theorem \cite{Beasley:2003fx}, which could zero out the superpotential. 
Recently it was shown in \cite{Buchbinder:2016rmw,Buchbinder:2017azb}, that for a complete  intersection Calabi-Yau  which has $h^{1,1}$ larger than the $h^{1,1}$ of its ambient space, such as the Schoen $\hets$, the Beasley-Witten vanishing criterion can be evaded.\footnote{ An explicit construction of homologically inequivalent curves was {given} in \cite{Buchbinder:2017azb}, and these are expected to contribute to the superpotential if the corresponding prefactors are non-vanishing.}

From each of the rational elliptic surfaces $dP_9$, there is an $E_8$ lattice worth of sections -- and we will provide a detailed derivation of this lattice using string junctions in section \ref{sec:StringJunk}. Notice also that the genus-zero topological string partition function for the A-model on this manifold has been computed \cite{Saito} and it indeed equals a product of two $E_8$ theta-functions, which confirms the curve counting of \cite{Donagi:1996yf,Curio:1997rn}.

We are interested in heterotic duals of the infinite number
of non-perturbative superpotential corrections in F-theory, and will thus focus on heterotic
worldsheet instantons since the divisors $D_B$ of \cite{Donagi:1996yf}
are of the form $D_B = \pi_B^{-1}(C)$. The D3-brane (or M-theory dual M5-brane) instanton zero modes studied
in \cite{Donagi:1996yf} were counted by structure sheaf cohomology
\begin{equation}\label{1000}
h^i(D,\CO_D) = (1,0,0,0),
\end{equation}
and in particular zero modes from the $3$-$7$ sector were not studied because $\Fth$
is smooth and there are no non-trivial seven-branes. It is assumed there and in
this work that potential zero-modes from instanton intersections with the $I_1$
locus are absent; to our knowledge, this issue has received relatively little
attention in the literature. 

Instead, we are interested in the heterotic worldsheet instanton zero modes
that are the duals of $h^{i}(D,\CO_D)$. They do not depend on the heterotic
vector bundle $\hbun$, which does appear in modes that are the duals of the
 $3$-$7$ modes, but instead only depend on the geometry of $C$ inside $\het$. 
 Since $\hbe$ is common to both the F-theory and heterotic compactification,
 it is useful to instead express the zero modes in terms of $C$ and $\hbe$
 rather than $C$ and $\het$; see e.g.\ \cite{Cvetic:2012ts} for a derivation.
 In this situation the condition on zero modes for a superpotential
 correction is
\begin{equation}
(h^0(C,\CO_C),h^1(C,\CO_C),h^0(C,N_{C|\hbe}),h^1(C,N_{C|\hbe}))=(1,0,0,0).
\label{eqn:hetmodes}
\end{equation}
The modes associated with $h^{0}(C,\CO_C)$ contribute the $\int d^2 \theta$
required for a superpotential correction, and the others must vanish so as
to not have too many Fermi zero modes. These zero mode considerations put strong constraints on $C$. The
condition $h^1(C,\CO_C)=0$ implies that $C$ must be a $\bbP^1$. Then,
 since $\CO(-1)$ is the only line bundle on $\bbP^1$ whose
 cohomology vanishes, we deduce that $N_{C|\hbe}=\CO_{\bbP^1}(-1)$. 

 In summary, in the absence of additional physics that might lift zero modes,
 the condition for a heterotic worldsheet on $C$ in $\hbe$ to contribute
 to the superpotential is that it be a rigid holomorphic curve of genus $0$. This
 implies the equation \eqref{eqn:hetmodes}.

Applied to the Schoen threefold, we would like to identify the heterotic duals to the infinite
number of sections contributing to the F-theory superpotential. Recall the divisors in the Calabi-Yau fourfold $Y_{\rm DGW}$ in F-theory that contributed D3-instanton corrections were of the type (\ref{DivsF}), i.e., pull-backs of $\sigma \times \mathbb{P}^1$, where $\sigma$ is a section of $\widehat{dP_9}=\widehat{S}$. 
In the dual heterotic compactification, $\hbe=\widehat{dP_9}$,
and therefore the same curves $C$ whose pullback into the K3-fibration of $Y_{\rm DGW}$ are wrapped by
M5-branes in the M-theory / F-theory description may be wrapped by
heterotic worldsheet instantons. These are rational curves. 
To determine their normal bundle 
note that $\widehat{dP_9}$ can be embedded with bidegree $(3,1)$ in $\bbP^2\times \bbP^1$,
and from this description an adjunction calculation shows that $N_{C|\widehat{dP_9}}=\CO(-1)$.
Therefore, \eqref{eqn:hetmodes} holds and we have superpotential corrections from
heterotic worldsheets on each $C$. 

To compute the superpotential in heterotic string theory, we need to evaluate
\begin{equation}\label{eq:hetsuperpotential}
W = \sum_C G_C  \, \exp\left[2 \pi i \int_C \mathbb{J} \right] \; ,
\end{equation}
where $\mathbb{J} = B + i \, J $ is the complexified K\"ahler form, for rigid holomorphic curves $C$ in $\hets$. 
Denote the product of elliptic fibers 
\be
F = \mathbb{E} \times \widehat{\mathbb{E}}\,,\qquad \hbox{\qquad} [\mathbb{E}] = E \,,\quad [\widehat{\mathbb{E}}] = \widehat{E} \,,
\ee
and each section of this fibration gives rise to a rigid $\mathbb{P}^1$. 
Both $\mathbb{E}$ and $\widehat{\mathbb{E}}$ are fibered individually over the base to give rise to the rational elliptic surfaces $S$ and $\widehat{S}$, respectively.  
The sections of the elliptic fibrations on $S$ and $\widehat{S}$, and correspondingly on $\het$, are described by \eqref{eq:sigmaandgamma}. Sections of $\hets$ are 
hence given by combining two such sections and are 
\begin{equation}\label{eq:hetcurvesrigid}
\sigma_{\gamma,\widehat\gamma} = \sigma_\gamma\cdot\sigma_{\widehat\gamma} = (\gamma + \sigma_0 + n {E}) \cdot (\widehat\gamma + \widehat\sigma_0 + \widehat{n} \widehat{{E}}) \,.
\end{equation}
Note that this entails that the same divisor, $F$ which corresponds to fixing a point on the $\mathbb{P}^1$ base of $\het$ appears in the expressions for $\sigma_\gamma$ and $\sigma_{\hat\gamma}$. We now parameterize the {complexified} K\"ahler form {$\mathbb{J} = B + i \, J $} as
\begin{equation}
 \mathbb J = (\sigma_0 + F)\tau + (\widehat\sigma_0+ F)\widehat\tau + F z +  \sum_i \omega_i \alpha_i^* + \hat\omega_i \hat\alpha_i^{*}
\end{equation}
and evaluate \eqref{eq:hetsuperpotential}. Note that $F^2 = 0$ and $F \cdot \sigma\cdot \widehat{\sigma}=1$ for any pair of sections $\sigma$ and $\widehat\sigma$.
As all sections of the double elliptic fibration are related by automorphisms of $\het$, the coefficient of the different terms in \eqref{eq:hetsuperpotential} cannot depend on geometric moduli. However, it can in principle depend on bundle moduli, which mirrors the situation in F-theory. 
Again we parameterize
$\gamma = \sum_i m_i \alpha_i$ and $\widehat\gamma = \sum_i \hat{m}_i \hat\alpha_i$. 
With this we find\footnote{ We may think of any section $\sigma_{\gamma}$ as restricting to $\widehat{dP_9}$. The intersections $\hat\gamma\cdot \sum_i \hat\omega_i \hat\alpha_i^{*}$ which results in $\sum_i  \hat{m}_i \hat\omega_i$ as the $\hat\alpha_i^{*}$ were chosen to form a dual basis to the $\hat\alpha_i$.}
\begin{equation}
\begin{aligned}
 W & = \sum_{E_8 \times E_8 } G_{\gamma,\hat\gamma} \exp\left[2\pi i (\gamma + \sigma_0 + n {E}) \cdot 
 (\hat\gamma + \hat\sigma_0 + \hat{n} \widehat{{E}})\cdot  \mathbb J \right]\\
  &=\sum_{m,\hat{m} \in \mathbb{Z}^8 \times \mathbb{Z}^8  } G_{m,\hat{m}} \exp 2\pi i \left[z + n \tau + \hat{n}\hat\tau +  \sum_i m_i \omega_i +  \hat{m}_i \hat\omega_i\right] \,,
 \end{aligned}
\end{equation}
{where the dependence on $n $ and $\hat n$ is as in \eqref{eq:ndoesnotmatter}.}Under the assumption that the moduli-dependent prefactors $G_{m, \hat{m}}$ are universal: {$G_{m, \hat{m}}= G$ for all $m,\hat{m}$,} this equals 
\be
W= {G}\,\, e^{2 \pi i z}\, \Theta_{E_8}(\tau,{\omega})\,\, \Theta_{E_8}(\hat\tau,{\hat \omega})\,,
\ee
where the K\"ahler moduli again need an appropriately rescaled. 
This is not strictly speaking allowed. The space-time filling D3 branes on the F-theory side are mapped under the duality to heterotic NS5 that are wrapping the elliptic fiber of $\het$ \cite{Andreas:1997ce}. Depending on  their positions along the base $\hbe$, additional zero-modes can arise that lift the corresponding instanton contribution, which is the dual effect to Ganor-strings on the F-theory side. We shall see the counterpart of this effect on the M-theory side of the duality in section \ref{sec:G2Insts}.

In the remaining part of this section we are going to reproduce this result using a string junction picture for heterotic instantons.


\subsection{Heterotic Instantons from String Junctions}
\label{sec:StringJunk}

The heterotic world-sheet instanton contributions on the Schoen Calabi-Yau threefold were thus far discussed using the description of the Schoen in terms of a double-elliptic fibration. This description is particularly useful to identify the dual contributions to the DGW superpotential in F-theory. To map this, however, to M-theory on a TCS manifold, we need to identify the heterotic world-sheet instantons in the alternative description of the Schoen as an SYZ-fibration (see figure \ref{fig:DualityChain}). A particularly useful way to approach this is using `string junctions' -- by this we mean the relative homology cycles
associated with string junctions, which in this case will be related to cycles wrapped
by heterotic worldsheet instantons; see \cite{Gaberdiel:1997ud,DeWolfe:1998zf}  for early physics work on string junctions, \cite{Grassi:2013kha} for realizations and explicit calculations in Weierstrass models, based on a rigorous geometric and topological treatment \cite{Grassi:2014ffa}.

This particular approach may seem ill-advised in the context of an SYZ-fibration of the Calabi-Yau threefold, as the $T^2$-fibrations we are interested in are not elliptic in the complex structure inherited from $\hets$. However,  in a twisted connected-sum description of the Schoen Calabi-Yau \cite{Braun:2017uku} each of the building blocks can be locally given a complex structure, where two of the circles of the SYZ-fibration can be thought of as an elliptic curve. This allows us to construct the curves, which correspond to the sections of the $dP_9$ surfaces in the Schoen Calabi-Yau, by gluing together `thimbles' from each building block.  

\begin{figure}
\begin{center}
\includegraphics[height=7cm]{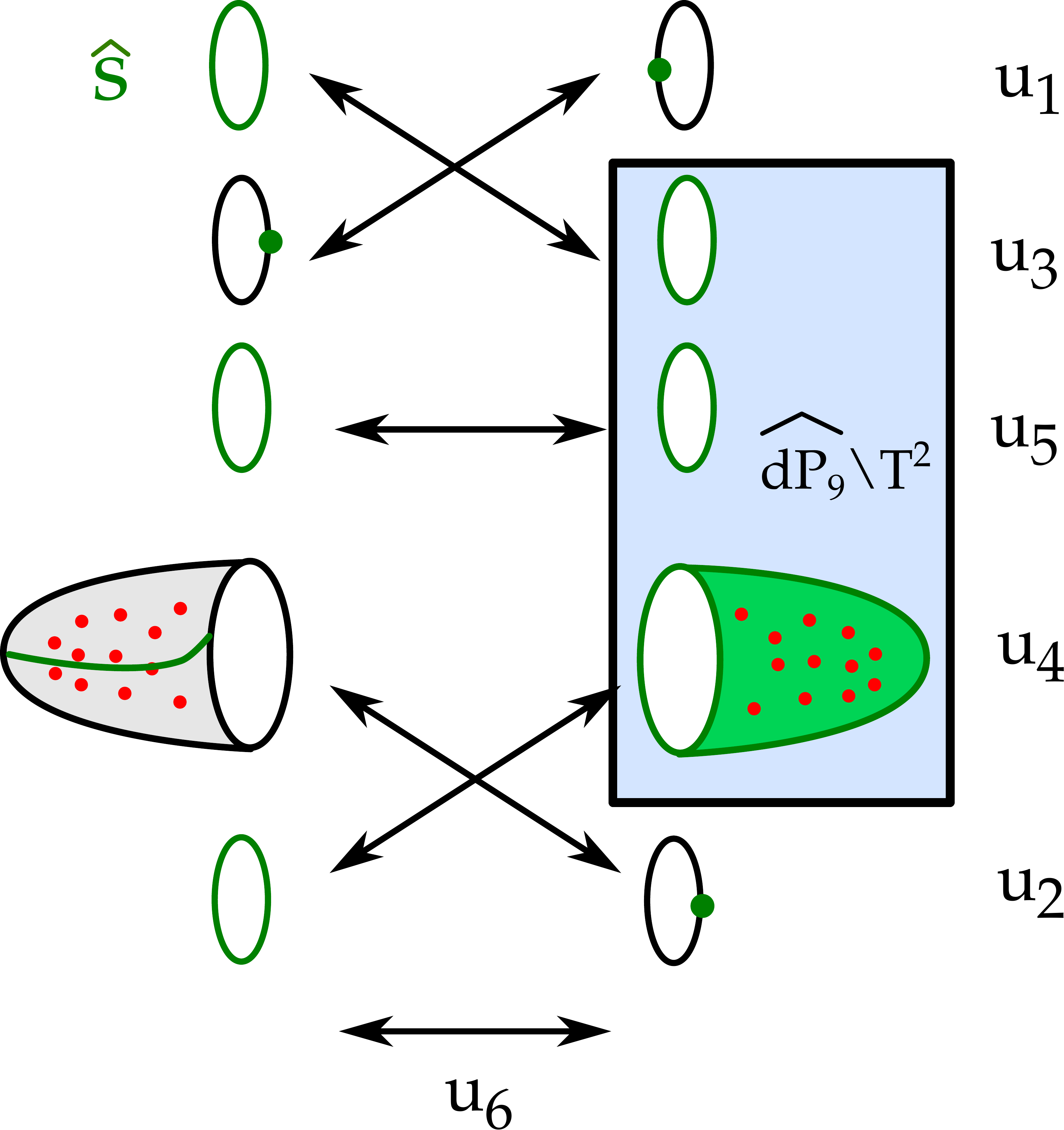} \qquad \qquad 
\includegraphics[height=7cm]{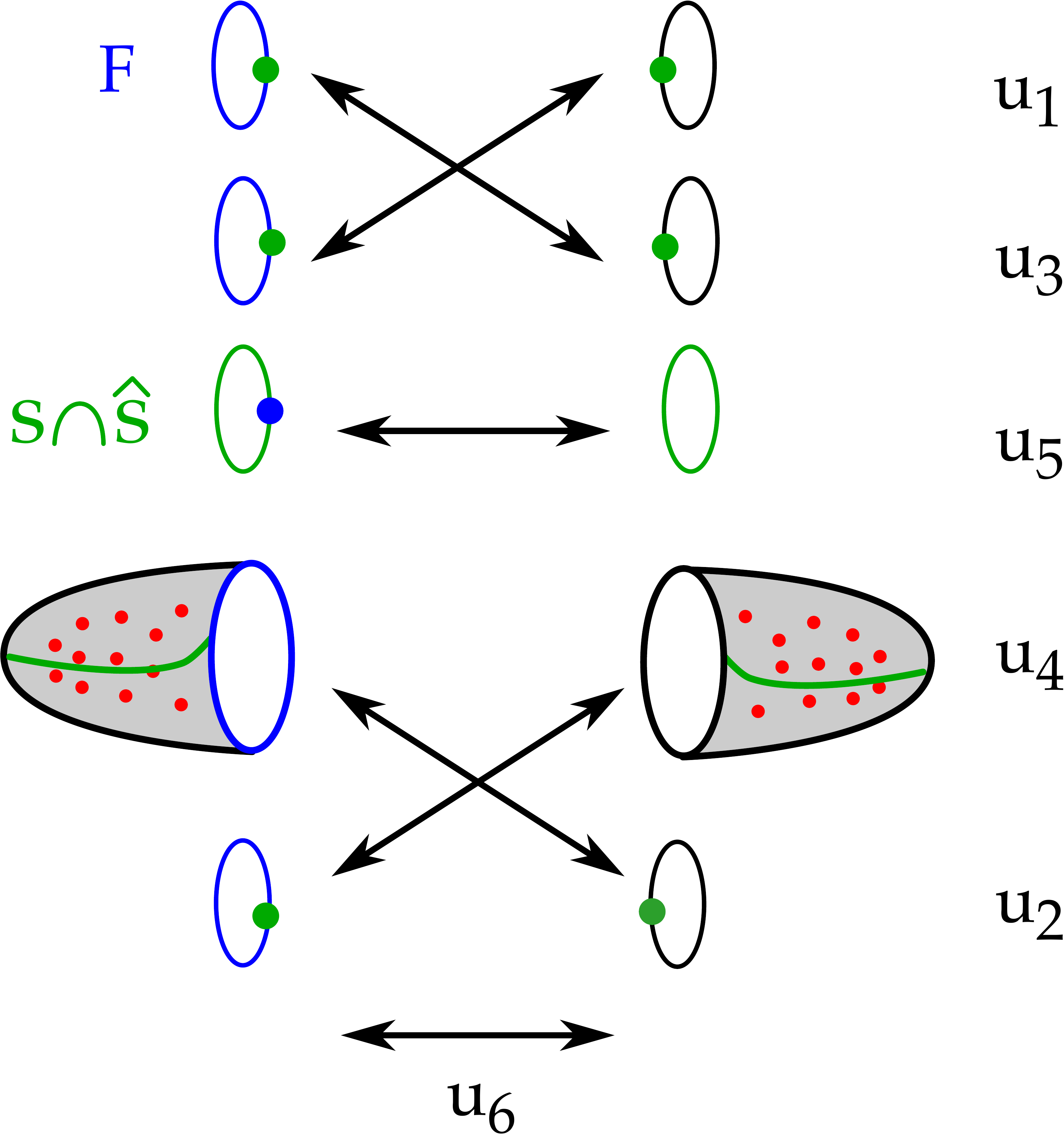}
\caption{The Schoen Calabi-Yau threefold as a connected sum. The figure on the left shows the surface {$\hat{S}$} in green. The figure on the right shows the divisor $F = \mathbb{E} \times \widehat{\mathbb{E}}$ in blue and the intersection $S \cap \hat{S}$ in green. {The coordinates are labeled by $u_1, \cdots, u_6$. }
\label{fig:good_fibrations}}
\end{center}
\end{figure}

First we recall the twisted-connected sum description of the Schoen Calabi-Yau -- the reader can find a more in depth description in \cite{Braun:2017uku}. The building blocks, denoted $M_\pm$ in figure \ref{fig:DualityChain} are $T^3$-fibrations over a base $\mathbb{P}^1 \times \bbS_{e\pm}^1$ with a single point on the $\mathbb{P}^1$ and the fiber over it removed. As one of the circles in the $T^3$-fiber, $\bbS_{s,\pm}^1$, undergoes no monodromies over the base, and furthermore the fibration is trivial over $\bbS_{e\pm}^1$, we can write $M_\pm = V_\pm \times \mathbb{S}^1_{s,\pm} \times \mathbb{S}^1_{e,\pm}$ with $V_\pm = dP_9\backslash T^2$, see figure \ref{fig:good_fibrations}. In the region $M_+\cap M_- = I \times \mathbb T^5$ the coordinates $u_i$ associated with the various $\bbS^1$ factors are
\begin{equation}
\begin{array}{lll}
u_1 &\leftrightarrow& \bbS_{s-}^1  \\
u_2 &\leftrightarrow& \bbS_{e-}^1  \\
u_3 &\leftrightarrow& \bbS_{s+}^1  \\
u_4 &\leftrightarrow& \bbS_{e+}^1  \, .
\end{array}
\end{equation}
Furthermore, we can identify the $dP_9$ and $\widehat{dP}_9$ appearing in the realization of $\hets$ as a fiber product $\hets= dP_9 \times_{\widehat{\mathbb{P}}^1} \widehat{dP_9}$ with the $V_\pm$ appearing in the SYZ realization of $\hets$ as 
\begin{equation}
\begin{array}{lllll}
S \,\setminus\, \mathbb{E} &\equiv& dP_9 \setminus \mathbb{E}& = & V_+ \\
\widehat{S} \,\setminus\, \widehat{\mathbb{E}}&\equiv& \widehat{dP}_9\setminus \widehat{\mathbb{E}} & = &V_- \,.
\end{array}
\end{equation}
The crucial idea is that the sections of the elliptic fibrations on $dP_9 $ and $\widehat{dP}_9$ (inherited from the complex structure of $\hets$) become string junctions in the elliptic fibration on $V_\pm$ inherited from the $T^2$ contained in the $T^3$-fiber of the SYZ-fibration. This should not come as a surprise, as hyper-K\"ahler rotations in general map algebraic to transcendental cycles.

The fiber of the elliptic fibration on $V_\pm$ degenerates at 12 points, $\points_i$, $i=1, \cdots, 10$ and $\points_\pm$, on the base which is a $\mathbb{P}^1$ with a point removed. The fiber above each of the points $\points_i$ or $\points_\pm$ is
an $I_1$, whereby a one-cycle in the elliptic fiber collapses. The two-cycles relevant for the string junctions will be constructed from paths on the base, connecting $\points_i$ together with the collapsed 1-cycles. Each building block has this behavior with monodromies on one side of the connected sum construction.

To construct the curves $\sigma_{\gamma, \hat{\gamma}}$ in the $\hets$ in this description, we first study a slightly simpler problem of the curves in the open $dP_9\setminus T^2$ and then glue the two halves together to obtain the curves $\sigma_{\gamma, \hat\gamma}$ (\ref{eq:hetcurvesrigid}) of $\hets$ in the string junction picture. 
For $dP_9$ these curves were already determined in the language of string-junctions in \cite{DeWolfe:1998pr}. 
An in depth construction of the junctions relevant here can be found in appendix \ref{app:StringJ}, where we determine all the vanishing cycles and explicitly compute each
topological detail of $dP_9$ string junctions presented in this section. Note the vanishing cycles of
appendix \ref{app:StringJ} are different from those of \cite{DeWolfe:1998pr} though both give rise to the same topological results we have presented. Henceforth we use the conventions of 
\cite{DeWolfe:1998pr}.

In figure \ref{fig:good_fibrations}, the rational elliptic surface $\widehat{S}$ is shown inside the Schoen Calabi-Yau (the surface ${S}$ is found by swapping the left and right of the figure). 
Note that {$\widehat S$} is presented as a $dP_9$ on the right hand side, and on the left (with a different induced complex structure) it becomes a $T^2$ fibered over a junction in the open $\mathbb{P}^1$ on the left, with asymptotic charge  $[1,0]$. The goal is now to associate to each section of the $dP_9$ such a string junction (or thimble) $\mathfrak{t}_\gamma$, and to recover the section $\sigma_{\gamma,0}$ by capping it off appropriately. These
$\sigma_{\gamma,0}$ are then turned into the four-cycles $\sigma_\gamma$ by 
taking a product with an appropriate $T^2$. We will discuss each of these steps in
turn.

\begin{figure}
\begin{center}
\includegraphics[width=12cm]{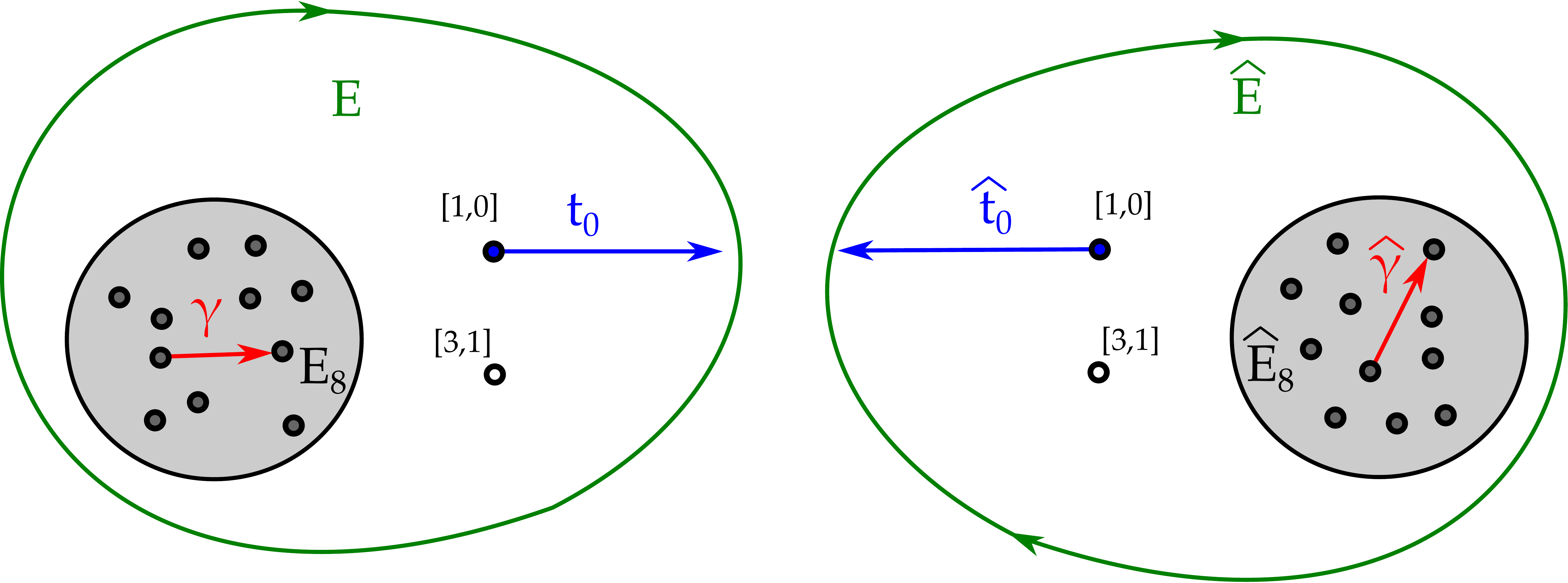}
\caption{The degenerations of the elliptic fibration over the building blocks {$V_\pm$}. The $E_8$ root lattice can be found from junctions between $10$ of the $12$ points and the remaining two can be chosen to have charges $[1,0]$ and $[3,1]$. The figure on the right shows the junctions associated with the terms in \eqref{eq:biffbangpow}. \label{fig:12isthenumber}}
\end{center}
\end{figure}

First we would like to associate a string junction $\mathfrak{t_\gamma}$ with each section
$\sigma_{\gamma,0}$.
Of the 12 degeneration points of the $T^2$ fiber, 10 realize the $E_8$ roots, whereas the remaining two correspond to asymptotic $[p,q]$ charges $[1,0]$ and $[3,1]$, respectively, see figure \ref{fig:12isthenumber}. 
Recall the sections of the rational elliptic surfaces take the form 
\be\label{eq:biffbangpow}
\sigma_{\gamma,0}= \gamma + \sigma_0  + n E \,,
\ee
with $\gamma^2=-2 n$ and a choice of $dP_9$ zero-section $\sigma_0$ (which is $\sigma_{0,0}$ in
the threefold) and fiber class $E$ of the $dP_9$-surface $S$, satisfying \eqref{SectionCond}. 
Each is a topological two-sphere that may be obtained (in a way described momentarily) from a junction representation
of the same object as
\begin{equation}
\mathfrak{t}_\gamma = \gamma + \mathfrak{t}_0 + n E.
\end{equation}
In terms of string junctions, $\gamma$ connects points $\points_i \rightarrow \points_j$, $i,j=1,\cdots, 10$ and gives rise to the $E_8$ lattice. The thimble $\mathfrak{t}_0$
with asymptotic charge $[1,0]$ may be capped off  into the zero-section $\sigma_0$, and the fiber $E$ encircles all nodes, see figure \ref{fig:12isthenumber}. The intersections of these
junction representations are
\begin{equation}
E^2=\gamma\cdot E = \gamma \cdot \sigma_0 = 0 \qquad \sigma_0 \cdot E = 1 \qquad \sigma_0^2=-1 \qquad \gamma^2 = -2n \,,
\end{equation}
as explicitly computed in appendix \ref{app:StringJ}, and as may be deduced from the figure.
These intersections ensure that the string junction satisfies $\mathfrak{t}_\gamma^2=-1$.

Since $\mathfrak{t}_\gamma$ has asymptotic charge, inherited entirely from
$\mathfrak{t}_0$, it has a boundary and cannot be a section $\sigma_{\gamma,0}$. However, it
may be capped off with a thimble from the other building block, specifically the
base of its open $dP_9$, which removes its boundary and preserves its self intersection
since this capping-off thimble has self-intersection $0$. This capping off is what 
allows us to associate a string junction with asymptotic charge $[1,0]$ (i.e., $\mathfrak{t}_0$) with
the zero section $\sigma_0$.

We have hence seen that the sections $\sigma_{\gamma,0}$ of {$V_+$} may be obtained by capping off
string junctions $\mathfrak{t}_\gamma$ on {$V_+$} with asymptotic charge $[1,0]$ and self-intersection $-1$. These string junctions are connected to the $12$ points of degeneration of the elliptic fibration on {$V_+$}, shown in figure \ref{fig:12isthenumber}. This figure also shows the junctions corresponding to elements $\gamma$ in the $E_8$ lattice, the asymptotic charge $[1,0]$ junction that becomes the zero-section after capping off, and the fiber $E$.
Note that the product of the monodromies associated with all of these $12$ degeneration points is trivial, so that the cycle $E$, which is identified with the restriction of $F$ to {$V_+$} by glancing at figure \ref{fig:good_fibrations}, exists.  

We can now represent any of the cycles \eqref{eq:biffbangpow} by a reducible combination of the cycles shown in figure \ref{fig:12isthenumber}. By moving the torus $E$ across the degeneration points via so-called Hanany-Witten moves, as we show in appendix \ref{app:StringJ}, it may be represented as a junction and the $\mathfrak{t}_\gamma$ become smooth two-cycles. 
Alternatively, as a further consistency check of the existence of these junctions $\mathfrak{t}_\gamma$
with $\mathfrak{t}_\gamma^2=-1$, we can appeal to the analysis of \cite{DeWolfe:1998pr}. There the self-intersection of any junction $J$ with asymptotic charge $[p,q]$ on a $dP_9$ has been determined as
\begin{equation}\label{eq:zwdint}
J^2 = \gamma^2 -2 n_1 k_1 -2 n_2 k_2  -( k_1^2 +k_2^2 +k_1 k_2)\,,
\end{equation}
where $\gamma$ is the part of the junction in $E_8$ and $n_1$ and $n_2$ count the number of prongs on the points with charges $[1,0]$ and $[3,1]$. The asymptotic charge $[p,q]$ of such a junction is related to $k_1$ and $k_2$ by
\begin{equation}
 k_1 = -q \hspace{1cm} k_2 = 3q-p \, .
\end{equation}
For the junctions $\mathfrak{t}_\gamma$ we are interested in, we have $p=1$ and $q=0$, so that $k_1=0$ and $k_2 = -1$. Then from \eqref{eq:zwdint} we have
\begin{equation}
\mathfrak{t}_\gamma^2 = \gamma^2 + 2 n_2  - 1 \,,
\end{equation}
which forces the choice $2n_2 = -\gamma^2$ in order to obtain $\mathfrak{t}_\gamma^2 = -1$.

We apply the construction of the sections of $dP_9$ to the {$V_\pm$} inside each building block $M_\pm$, which gives rise to thimbles $\mathfrak{t}_\gamma$ 
(or $\mathfrak{t}_{\hat \gamma}$ from the other building block) that may be
capped off to form $\sigma_{\gamma,0}$ (or $\sigma_{0,\hat \gamma}$). These may be promoted into four-cycles
via
\be
\sigma_\gamma = \sigma_{\gamma,0} \times T^2_{{u_1, u_2}}\,,\qquad 
\hat\sigma_{\hat{\gamma}} = \sigma_{0,\hat{\gamma}} \times T^2_{{ u_3, u_4}} \,,
\ee
where the coordinates indicate the transverse $T^2$ to the rational elliptic surfaces $S$ and $\hat{S}$, respectively, as shown in figure \ref{fig:good_fibrations}.
These are how the four-cycles $\sigma_\gamma$ and $\sigma_{\hat \gamma}$ studied in
the double elliptic fibration description of the Schoen
arise in its SYZ description. From them we form the usual two-cycles
by intersection of the two divisors
\begin{equation}
\sigma_{\gamma, \hat \gamma } = \sigma_\gamma \cdot \sigma_{\hat \gamma},
\end{equation}
which may be wrapped by heterotic worldsheet instantons that correct the superpotential.
Note that the fact that sections of the elliptic fibration on $S$ and $\widehat{S}$ square to $-1$ implies that $\hat{S} \cdot \hat{S} = \sigma_0 \cdot \sigma_0 = -\hat{E}$. We can compute the self-intersection from the presentation of figure \ref{fig:good_fibrations} by noting that a homologous cycle to $\hat{S}$ can be completely displaced along the product $\mathbb{S}^1$s on the right hand side, so that the self-intersection comes purely from the self-intersection of the string junction, which is consistent with the capping off of the string junction not changing
its self-intersection. The locus of self-intersection is geometrically given by the two product circles on the right side, which is nothing but $\hat{\mathbb{E}}$.

\section{Instantons from Associatives in TCS $G_2$-manifolds}
\label{sec:G2Insts}

In this section, we use the duality between heterotic string theory on $\hets$ and M-Theory on $J$ to lift the rigid holomorphic $\mathbb{P}^1$s on $\hets$, that give rise to world-sheet instantons, to associative three-cycles on $J$. As we consider heterotic models dual to F-Theory on $Y_4$, the vector bundle $\hbun$ on the heterotic side is chosen such that it completely breaks the $E_8 \times E_8$ gauge symmetry and satisfies $ch_2(\hbun) = 12\widehat{\mathbb{E}}$. As explained in section \ref{sec:FHG}, this means that the dual geometry on the M-Theory side is a TCS $G_2$-manifold $J$ glued from the two building blocks $Z_\pm$ with 
\begin{equation}
 \begin{aligned}
  N_+ &= U_2 \\
  N_- &= U_3 \oplus (-E_8) \oplus (-E_8) \, .
 \end{aligned}
\end{equation}
We will hence be interested in how the contributions to the superpotential discussed in the last section show up in the geometry of $J$. As $J$ is formed as a TCS $G_2$-manifold, we start by explaining the geometry of the building blocks $Z_\pm$ in  detail.

\subsection{The Geometry of  $Z_-$ and $S_-$}

The threefold $Z_-$ is described algebraically by 
\be \label{eq:weierstrassZ-}
y^2 = x^3 + f_{4,8}(\hat{z},z) x + g_{6,12}(\hat{z},z) \; ,
\ee
where $f$ and $g$ are homogeneous polynomials of the indicated degrees in the homogeneous coordinates $[z_1:z_2]$ and $[\hat{z}_1:\hat{z}_2]$ on $\mathbb{P}^1\times \hat{\mathbb{P}}^1$. In particular, $[z_1:z_2]$ are homogeneous coordinates on the $\mathbb{P}^1$ base of the elliptically fibered K3 surface $S_-$, and $[\hat{z}_1:\hat{z}_2]$ are homogeneous coordinates on the base $\hat{\mathbb{P}}^1$ of the K3-fibration on $Z_-$.
The polynomials $f_{4,8}$ and $g_{6,12}$ are furthermore chosen such that
\be \label{eq:weierstrassfg-}
\begin{split}
f_{4,8} &= \alpha (\hat{z}) z_1^4 z_2^4 \\
g_{6,12} &= \delta (\hat{z}) z_1^5 z_2^7 + \beta (\hat{z}) z_1^6 z_2^6 + \delta'(\hat{z}) z_1^7 z_1^5 \; .
\end{split}
\ee
A generic fiber $S_-$ is found by fixing the coordinate $\hat{z}$ to a generic value. Of course, the geometry described above is fairly singular, and we need to resolve the singularities to arrive at a smooth building block. This is done by blowing up the two $E_8$ singularities ($II^*$ fibers), as well as the twelve points $\delta\delta'=0$ over which there is a remaining point-like singularity\footnote{ This singularity is of type $\tilde{E}_8$ with a local model $x_1^2+x_2^3 + x_3^6+x_4^6=0$. It is an isolated threefold singularity which has a crepant blow-up with exceptional divisor $dP_9$. After the resolution, the K3 fibers over $\delta\delta'=0$ become reducible with two components each isomorphic to $dP_9$, one of which is the exceptional divisor.}. As these resolutions do not alter the transcendental cycles of $S_-$ nor the monodromies acting on them, we leave this resolution implicit. After this resolution is performed, the exceptional cycles of the two $E_8$s together with the section and fiber of the elliptic fibration generate the lattice $N_- =  U_3 \oplus (-E_8) \oplus (-E_8)$, {whose orthogonal complement is} $T_- = U_1 \oplus U_2$. The geometry of $S_-$ together with its monodromy group has been previously discussed in some detail in \cite{LopesCardoso:1996hq,McOrist:2010jw,Braun:2013yla,Malmendier:2014uka,Garcia-Etxebarria:2016ibz}.

\begin{figure}[tb]
\centering
\includegraphics[width=0.3\textwidth]{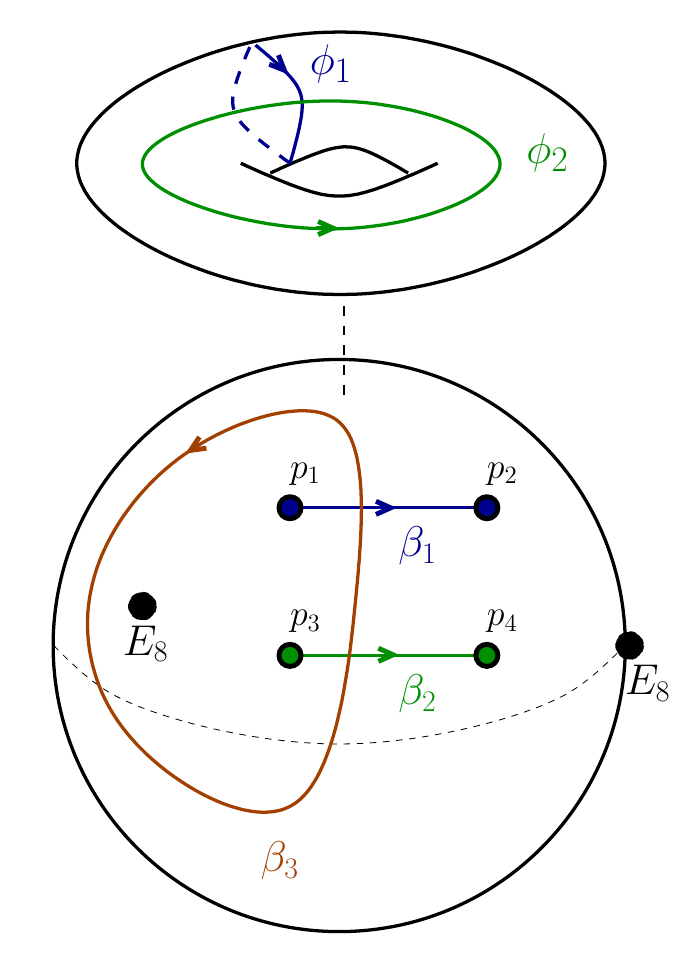}
\caption{A cartoon of the elliptically fibered K3 surface, with the $E_8$ singular points at $z_{1,2}=0$. $\beta_3$ is a one-cycle on the base $\mathbb{P}^1$, which combine with the two one-cycles $\phi_{1,2}$ of the elliptic fiber to form two genus-one two-cycles $e^1, e^2$. The one-cycles $\phi_{1}$ ($\phi_2$) in the fiber shrink to zero volume at the points $p_{1,2}$ ($p_{3,4}$). Thus, we get  two $-2$ curves by fibering $\phi_{1,2}$ over the intervals $\beta_{1,2}$ on the base $\mathbb{P}^1$. We denote these curves by $e_1-e^1$ and $e_2-e^2$.  }\label{fig:k3cycles}
\end{figure}

The geometry of $S_-$ can be easily understood by exploiting its elliptic fibration.
The discriminant of the elliptic fibration on $S_-$ follows from \eqref{eq:weierstrassfg-} and can be written as (in a patch where $z_2=1, z=z_1/z_2$):
\be
P(z,\hat{z}) = z^{10} P_4 \; , \hspace{.5cm} P_4 =  4 \alpha^3 z^2 + 27 ( \delta'z^2 + \beta z + \delta)^2  \,,
\ee
where we have suppressed the $\hat{z}$ dependence of $\alpha, \beta, \delta$ and $\delta'$. Besides the two $II^*$ fibers, there are four special points $p_1,\cdots, p_4$ above which the elliptic fiber degenerates inducing monodromy maps with $(p,q)$-charges $Q_i$, see figure \ref{fig:k3cycles}. It can be shown \cite{DeWolfe:1998pr} that these pairwise have the same $SL(2,\mathbb{Z})$ monodromy acting on the elliptic fiber, which we may choose as
\be
\ba
Q_1 = Q_2 = &[1,0] \\
Q_3 = Q_4 = &[3,1] \,.
\ea
\ee
Furthermore, the monodromy around e.g.\ $p_1$, $p_3$ together with one of the $E_8$ stacks, i.e., around the loop $\beta_3$ in figure \ref{fig:k3cycles} is trivial. This allows us to construct the lattice $T=U_1 \oplus U_2$ of $S$ as follows. First, we may take any one of the $\mathbb{S}^1$s in the elliptic fiber over the loop $\beta_3$ to find a non-trivial two-cycle. This gives two independent cycles with the topology of a two-torus which each have self-intersection $0$ and do not mutually intersect. Let us denote them by $e^1$ and $e^2$. Furthermore, there are two-cycles with the topology of a two-sphere stretched between $p_1$ and $p_2$, as well as $p_3$ and $p_4$, respectively. These each have self-intersection $-2$, do not mutually intersect and each meet one of the two cycles $e^1$ and $e^2$ in a single point, so that we can associate them with $e^1 - e_1$ and  $e^2 - e_2$. Altogether, these four cycles hence span the lattice $U_1 \oplus U_2$ with inner {product matrix} 
\be \label{eq:cycleint}
e_i \cdot e^j = \delta_i{}^j \; .
\ee
Figure \ref{fig:k3cycles} is a cartoon of $S_-$, which shows these two-cycles.

The building block $Z_-$ is a holomorphic fibration of $S_-$ over $\hat{\mathbb{P}}^1$, so that the complex structure of $S_-$ varies as we move along $\hat{\mathbb{P}}^1$. The complex structure moduli space of $S_-$ is given by \cite{Aspinwall:1994rg,Aspinwall:1996mn}
\be \label{eq:monomonomonodromy}
\begin{split}
{\cal M}_{S_-}&= SO(2,2;\mathbb{Z}) \backslash O(2,2;\mathbb{R})/(O(2;\mathbb{R})\times O(2;\mathbb{R})) \\
&\approx (SL_{\tau}(2,\mathbb{Z}) \times  SL_{\sigma}(2,\mathbb{Z})  \ltimes \mathbb{Z}_{2})  \backslash \left( (SL_{\tau}(2,\mathbb{R}) /U(1)) \times  (SL_{\sigma}(2,\mathbb{R}) /U(1)) \right) \; .
\end{split} \ee
This space is spanned by two complex parameters $\tau$ and $\sigma$ and we may parameterize
\be\label{eq:omegaparam_ts}
\Omega^{(2,0)} = \tau e_1 + \sigma e^1 + e_2 - \tau \sigma e^2 \; .
\ee
It is straightforward to check that $\Omega^{(2,0)} \wedge \overline{\Omega^{(2,0)}} \sim \rm{Im}(\tau) \rm{Im}(\sigma) \neq 0$ and $\Omega^{(2,0)} \wedge \Omega^{(2,0)} =0$. 

The two modular parameters $\tau$ and $\sigma$ can be identified with complex structure $\tau$ of the dual elliptic curve $\mathbb{E}_h$ on the heterotic side and its complexified volume
\begin{equation}
 \sigma = \int_{{\mathbb{E}_h}} B + i J_h
\end{equation}
where $J_h$ is the K\"ahler form on $\hets$.

If follows from \eqref{eq:omegaparam_ts} that the parameters $\tau$ and $\sigma$ can be expressed as
\begin{equation}\label{eq:taufromomega}
 \tau =  \frac{\int_{e^1}\Omega^{2,0}}{\int_{e^2}\Omega^{2,0}} = -\frac{\int_{e_2}\Omega^{2,0}}{\int_{e_1}\Omega^{2,0}} \, ,
\end{equation}
and 
\begin{equation}
 \sigma =  \frac{\int_{e_1}\Omega^{2,0}}{\int_{e^2}\Omega^{2,0}} = -\frac{\int_{e_2}\Omega^{2,0}}{\int_{e^1}\Omega^{2,0}}\, .
\end{equation}
This allows us to determine the monodromy matrices related to any map in \eqref{eq:monomonomonodromy}. Consider the modular transformations associated to the $SL_{\tau}(2,\mathbb{Z})$ associated with $\tau$. From the action 
\begin{equation}
 \tau \rightarrow \frac{a\tau+b}{c\tau+d}
\end{equation}
of $SL_{\tau}(2,\mathbb{Z})$ on $\tau$, the action on $(e_1,e^1,e_2,e^2)^T$ is given by
\begin{equation}
M_\tau(a,b,c,d) =  \left(
\begin{array}{cccc}
 d & 0 & -c & 0 \\
 0 & a & 0 & b \\
 -b & 0 & a & 0 \\
 0 & c & 0 & d \\
\end{array}
\right)\, .
\end{equation}
Similarly, the action of $SL_{\sigma}(2,\mathbb{Z})$ is
\begin{equation}
M_\sigma(a,b,c,d) = \left(
\begin{array}{cccc}
 a & 0 & 0 & b \\
 0 & d & -c & 0 \\
 0 & -b & a & 0 \\
 c & 0 & 0 & d \\
\end{array}
\right)\, .
\end{equation}
Note that these matrices commute for any pair of elements $g \in SL_{\tau}(2,\mathbb{Z})$ and $g' \in SL_{\sigma}(2,\mathbb{Z})$, i.e.,  $[M_\tau(a,b,c,d),M_\sigma(a',b',c',d')]=0$. 
The relation between the parameters $\tau, \sigma$ and $\alpha, \beta, \delta, \delta'$ are \cite{LopesCardoso:1996hq, McOrist:2010jw, ElkiesKumar}
\be \label{eq:jrelations}
\begin{split}
-\frac{\alpha^3}{27 \delta \delta'} &= j(\tau) j(\sigma) \\
\frac{\beta^2}{4 \delta \delta'} &= (j(\tau)-1) (j(\sigma)-1) \,.
\end{split}
\ee
Modulo conjugation, the monodromies acting on $\tau$ and $\sigma$ are induced by the three special points in the fundamental domain of the $\tau$ (or $\sigma$) plane which are related to standard elements of $SL(2,\mathbb{Z})$: 
\begin{equation}
 \begin{array}{c|c|c}
  \tau & j(\tau) & \mbox{monodromy} \\
  \hline
  i & 1& T\\
i\infty & \sim e^{-2\pi i \tau}  & S\\
  e^{2\pi i/3} & 0 & ST 
 \end{array}
\end{equation}
The relations \eqref{eq:jrelations} can be (locally) solved to give 
\be  \label{eq:jrelations2}
j(\tau) = \frac{Q - \sqrt{\Delta_{S_-}}}{216 \delta \delta'} \, ,\qquad 
j(\sigma) = \frac{Q + \sqrt{\Delta_{S_-}}}{216 \delta \delta'}  
\ee
with 
\be
\begin{aligned}
Q =  4\alpha^3 + 27 \beta^2 - 108 \delta \delta' \, , &\hspace{.5cm}& \Delta_{S_-} = Q^2 + 1728 \alpha^3 \delta \delta' \; .
\end{aligned}
\ee

The K3 surface $S_-$ degenerates over $36 = 24+12$ points in the base $\hat{\mathbb{P}}^1$, which we will now describe. Any degeneration of the K3 surface $S_-$ will come from giving special locations to the points $p_i$, described by the vanishing of $P_4$. The discriminant of the polynomial $P_4$ is 
\begin{equation}
 \Delta(P_4) = \alpha^6 \delta^2 \delta'^2 \Delta_{S_-} \, .
\end{equation}
First note that the $P_4$ becomes a square, when $\alpha=0$, but there is no associated singularity of the K3 surface $S$. The elliptic fibration on $S$ just develops two fibers of type II when we are at this point in moduli space. This means e.g.\ the points $p_1$ and $p_3$ coincide, as do $p_2$ and $p_4$.

Whenever $\Delta_{S_-}=0$, which happens over $24$ points in the base $\hat{\mathbb{P}}^1$, the K3 fiber $S_-$ acquires an $A_1$ singularity. As $\Delta_{S_-}$ also appears as a factor in the discriminant of the polynomial $P_4$, two of the four special points in $z$ in the picture \ref{fig:k3cycles} come together whenever $\Delta_{S_-}=0$. 
From \eqref{eq:jrelations2}, it is clear that this happens when $j(\tau)=j(\sigma)$, so that $\tau = \sigma$ modulo $SL(2,\mathbb{Z})$. For a fixed point where $\Delta_{S_-}=0$, we may then pick a basis where $\tau = \sigma$, which means that the two-cycle $\gamma = e_1-e^1$ is orthogonal to $\Omega^{(2,0)}$: $\gamma \cdot \Omega^{(2,0)}=0$. Hence, this cycle vanishes as we approach the $\tau = \sigma$ locus, and, by the Picard-Lefschetz formula, we have that upon transport around this point, the two-cycles $(e_1, e^1, e_2, e^2)^T$ transform as
\be\label{eq:monoswape1e1}
M_{\tau \leftrightarrow \sigma} =
\begin{pmatrix}
0 & 1 & 0 & 0\\
1 & 0 & 0 & 0\\
0 & 0 & 1 & 0\\
0 & 0 & 0 & 1\\
\end{pmatrix} \; ,
\ee
i.e., $e_1 \leftrightarrow e^1$, which swaps the roles of $\tau$ and $\sigma$. These monodromies hence correspond to T-duality on the heterotic side. Note that by \eqref{eq:cycleint} $\gamma$ is a $-2$ curve, and hence topologically a two-sphere. In figure \ref{fig:k3cycles}, the shrinking of $\gamma$ corresponds to merging the points $p_1$ and $p_2$. However, note that  the identification of $\gamma$ depends on the basis choice we made for $T$, which is only determined up to modular transforms. Having made this basis choice at one singular point $\hat{z}_*$ in $\cal{B}$, we are not free to arbitrary change basis at another point. In stead, we must account for how the basis  transforms under transport along a path that connects these points. Thus, different $-2$ curves (that are related by modular transformations) will vanish at the 24 points in $\widehat{\mathbb{P}}^1$ where $\Delta_{S_-}=0$. 

The remaining degenerations of $S_-$ correspond to the $12$ points where $\delta \delta'=0$. Geometrically, two of the four points $p_i$ in figure \ref{fig:k3cycles} move on top of the loci of the $II^*$ fibers, so that the elliptic fibration on the K3 $S_-$ surface becomes non-minimal. After a resolution, we get a reducible K3 fiber over such points with an associated monodromy $T_\sigma$ modulo conjugation. These will act (modulo conjugation) as
\begin{equation}
T_\sigma : \sigma \rightarrow \sigma + 1\, , 
\end{equation}
so that the heterotic $B$-field is shifted by one unit. The loci $\delta\delta'=0$ are hence identified with the locations of NS5-branes in the dual heterotic theory.

The K3 surface $S_-$ enjoys a particularly nice limit in which the monodromies swapping $\tau$ and $\sigma$ are completely absent \cite{McOrist:2010jw,Garcia-Etxebarria:2016ibz}. This is achieved by turning $\Delta_{S_-}$ into a perfect square (and is related to an underlying Shioda-Inose structure). Let us set 
\begin{equation}\label{eq:shioda_inose_magic}
 \begin{aligned}
\alpha & = -3 \phi_\tau \phi_\sigma \\
\beta &=-\tfrac{27}{2} \gamma_\tau \gamma_\sigma \\
\delta \delta'&= \tfrac14 (4 \phi_\tau^3 + 27 \gamma_\tau^2)\,\, \tfrac14 (4 \phi_\sigma^3 + 27 \gamma_\sigma^2)
 \end{aligned}
\end{equation}
for some suitable polynomials $\phi_\tau,\gamma_\tau$ and $\phi_\sigma,\gamma_\sigma$. In this parameterization
\begin{equation}
 \Delta_{S_-} = 3^{12}(\phi_\tau^3 \gamma_\sigma^2 - \phi_\sigma^3 \gamma_\tau^2)^2
\end{equation}
and \eqref{eq:jrelations2} becomes
\begin{equation}
j(\tau)   = \frac{4}{27} \frac{ \phi_\tau^3 }{4 \phi_\tau^3 + 27 \gamma_\tau^2}\,,\qquad 
j(\sigma) = \frac{4}{27} \frac{\phi_\sigma^3}{4 \phi_\sigma^3 + 27 \gamma_\sigma^2}\,.
\end{equation}

\subsection{The Geometry of $Z_-$ in the Degeneration Limit}

Under the duality to M-Theory, the geometric regime of heterotic string theory is mapped to a specific limit of $S_-$, which corresponds to $\alpha,\beta \rightarrow \infty$, while keeping $\alpha^3/\beta^2$ fixed \cite{Morrison:1996na}. We hence rescale $\alpha$ by $\lambda^2$ and $\beta$ by $\lambda^3$ and let $\lambda \rightarrow \infty$. Equivalently, one may apply this limit as $\delta\delta' \rightarrow 0$. 

In this limit, the base $S^2$ of the K3 surface shown in figure \ref{fig:k3cycles} grows very long and the elliptic fiber becomes constant over the middle region in between the locations of the two $II^*$ fibers. The elliptic curve in this middle region is then identified with the geometry of the dual heterotic compactification. Using the basis of cycles constructed above, the complex structure (or rather, the ratio of the two radii of a basis of one-cycles) of the dual heterotic torus are hence given by
\begin{equation}\label{eq:liftsl2z}
 \tau_{het} =  \tau = \frac{\int_{e^1}\Omega^{2,0}}{\int_{e^2}\Omega^{2,0}}  \, . 
\end{equation}
The monodromies of the heterotic torus, which give rise to the geometry on the heterotic side, are identified with the subgroup $SL_{\tau}(2,\mathbb{Z})$ of the monodromy group in \eqref{eq:monomonomonodromy}. 

\begin{figure}
\begin{center}
  \scalebox{.5}{ 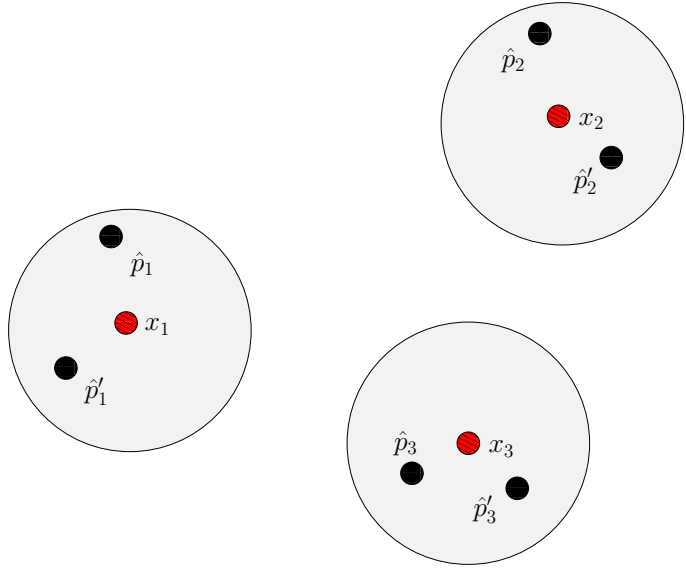 }
\caption{In the base $\hat{\mathbb{P}}^1$ of the K3-fibration on $Z_-$, the loci $\{\hat{p}_i,\hat{p}_i'\}$ defined by $\Delta_{S_-} = 0$ pairwise come close to the points $\{x_i\}$ defined by $\Delta_{het}=0$ in the limit $\lambda \rightarrow \infty$. Avoiding the shaded regions, the size of which goes like $\lambda^{-3}$, monodromies corresponding to T-dualities on the heterotic side are avoided. 
\label{fig:delta_het_S}}
\end{center}
\end{figure}

In the degeneration limit $\lambda \rightarrow \infty$, the $24$ monodromy points corresponding to swapping $\tau$ and $\sigma$, which are located at
\be
\Delta_{S_-} \sim 4 \alpha^3 + 27 \beta^2 + \mathcal{O}(\lambda^{-6}) = 0\, ,
\ee
are confined to small regions around the loci
\be
\Delta_{het} = 4 \alpha^3 + 27 \beta^2 = 0 \, ,
\ee
which are the $12$ degeneration points of the dual heterotic $\mathbb{E}_h$, see figure \ref{fig:delta_het_S}. The monodromy points corresponding to the $\mathbb{Z}_2$ exchanging $\sigma$ and $\tau$ in \eqref{eq:monomonomonodromy} hence come pairwise together, so that, apart from small regions, we can globally distinguish $\tau$ and $\sigma$.

Cutting out these small regions, $\tau$ and $\sigma$ become globally well-defined and can be written as
\begin{equation}\label{eq:tausigmaexpansion}
\begin{aligned}
j(\tau) \sim \frac{\alpha^3}{4 \alpha^3 + 27 \beta^2} + \mathcal{O}(\lambda^{-6})&\hspace{1cm}&
j(\sigma) \sim \frac{\lambda^6 (4 \alpha^3 + 27 \beta^2)^2  + 27\beta^2 \delta \delta'}{\delta \delta'(4 \alpha^3 + 27 \beta^2)} + \mathcal{O}(\lambda^{-6})
\end{aligned}
\end{equation}
by expanding \eqref{eq:jrelations2}.

When we move along the base $\hat{\mathbb{P}}^1$ we encounter various monodromies which are related to special points of the functions $j(\tau)$ and $j(\sigma)$. In the following, we shall work out these monodromies in the limit $\lambda \rightarrow \infty$. First note that whenever $\delta\delta'=0$, $j(\sigma)\rightarrow \infty$, so that we encounter (a conjugate of) the map $T_\sigma$. As before, these points are identified as the locations of the $12$ NS5-branes in the dual heterotic geometry. 

Let us now examine the $24$ points given by $\Delta_{S_-}=0$. We can group those $24$ points into $12$ pairs $\{\hat{p}_i,\hat{p}_i'\}, i=1\cdots12$ 
which merge pairwise in the vicinity of $\Delta_{het}=0$. Picking an arbitrary such point, $\hat{p}_1$ say, we may choose a basis in which the points $p_1$ and $p_2$ come together in the K3 fiber $S_-$ over $\hat{p}_1$, so that the cycle $e_1-e^1$ is collapsing there. As discussed before, this induces a monodromy given by acting with the matrix $M_{\hat{p}_1}$ \eqref{eq:monoswape1e1} on $(e_1,e^1,e_2,e^2)^T$. In the limit $\lambda \rightarrow \infty$, the point $\hat{p}_1'$ which approaches $\hat{p}_1$ induces the same monodromy up to conjugation. A careful analysis of the behavior of the points $p_i$ in the K3 fiber reveals that the points $p_1$ and $p_2$ coalesce over $\hat{p}_1'$ as well, albeit along a different path in the base $\mathbb{P}^1$ of the K3 surface $S_-$. This path is such that the cycle $e_1-e^1-e^2$ collapses\footnote{ Even without a detailed analysis, this can be argued for by noting that the product of the two associated Picard-Lefschetz transformation must be of infinite order and that we are furthermore free to exploit automorphisms of $U_1\oplus U_2$.}, which results in the Picard-Lefschetz monodromy associated with the matrix
\begin{equation}
M_{\hat{p}_1'} = \begin{pmatrix}
 0 & 1 & 0 & 1 \\
 1 & 0 & 0 & -1 \\
 -1 & 1 & 1 & 1 \\
 0 & 0 & 0 & 1 
\end{pmatrix} \; .
\end{equation}
Together, these two points hence generate a monodromy map 
\begin{equation}\label{eq:monoDelta_E}
M_{\hat{p}_1} M_{\hat{p}_1'} = \begin{pmatrix}
   1 & 0 & 0 & -1 \\
 0 & 1 & 0 & 1 \\
 -1 & 1 & 1 & 1 \\
 0 & 0 & 0 & 1 
 \end{pmatrix}\, .
\end{equation}

This result can be derived in an alternative way by exploiting the parameterization \eqref{eq:shioda_inose_magic} in the degeneration limit $\lambda\rightarrow \infty$. As discussed in \cite{Garcia-Etxebarria:2016ibz}, this leads to monodromies of the form 
\begin{equation}\label{eq:matrixofdoom}
M(a,b,c,d) \equiv M_\tau(a,b,c,d) \,M^{-1}_\sigma (a,b,c,d) = \left(
\begin{array}{cccc}
 d^2 & -b c & -c d & -b d \\
 -b c & a^2 & a c & a b \\
 -b d & a b & a d & b^2 \\
 -c d & a c & c^2 & a d \\
\end{array}
\right)
\end{equation}
for loops encircling any of the twelve points $\Delta_{het} = 0$, consistent with the fact that both $\tau$ and $\sigma$ degenerate at $4\alpha^3 + 27 \beta^2=0$. In particular, note that this reproduces \eqref{eq:monoDelta_E} for the T monodromy (i.e., for $a=b=d=1$ and $c=0$).

Let us summarize our result. In the degeneration limit $\lambda \rightarrow \infty$, the only monodromies acting on $S_-$ are 
\begin{equation}
 \begin{array}{c|c}
 \mbox{locus} & \mbox{monodromy} \\
 \hline
  4 \alpha^3 + 27 \beta^2 = 0 & T_\tau T^{-1}_\sigma \\
  \delta\delta'=0 & T_\sigma
 \end{array}
\end{equation}
up to conjugation. Note that something interesting has happened here. Although we only encounter a monodromy map of order two (corresponding to T-duality) for each of the $24$ points $\{\hat{p}_i,\hat{p}_i'\}$ given by $\Delta_{S_-}=0$, these points pairwise coalesce to generate monodromies of infinite order in the limit $\lambda\rightarrow \infty$. Of course, this can only happen because the points which come together pairwise correspond to different $\mathbb{Z}_2$ subgroups of \eqref{eq:monomonomonodromy}. Equivalently, there are different vanishing cycles at $\hat{p}_i$ and $\hat{p}_i'$ and we chose a basis where these are $e_1-e^1$ and $e_1-e^1-e^2$ for any given $i$. 

We are now ready to discuss the action of these monodromies on the $-2$ curves in $T_- = U_1 \oplus U_2$ and the corresponding lift of cycles from the two halves of the $\hets$. In the $dP_9$ studied in section \ref{sec:StringJunk}, the $SL_{\tau_{het}}(2,\mathbb{Z})$  monodromies are induced by the vanishing of the $(p,q)$ cycle of the fiber $T^2$, which induces a monodromy with 
\begin{equation}
\left(
\begin{array}{cc}
 a & b \\
 c & d \\
\end{array}
\right) = \left(
\begin{array}{cc}
 1-p q & p^2 \\
 -q^2 & 1 + pq \\
\end{array}
\right)\, .
\end{equation}
The $SL_{\tau_{het}}(2,\mathbb{Z})$ orbit of the cycle $(1,0)$ is hence given by
\begin{equation}
 (1 - p q, -q^2)\, .
\end{equation}
In the lift to M-Theory, the corresponding objects are as follows: the vanishing cycles at an arbitrary pair of monodromy points $\hat{p}_1$ and $\hat{p}_1'$ close to a degeneration point $\points_1$ ({\it cf.} figure \ref{fig:delta_het_S}) are chosen to correspond to the $(p,q)=(1,0)$ vanishing cycle on the heterotic side and are given by
\begin{equation}\label{eq:my_bicycles_have_vanished} 
\begin{aligned}
v_{(1,0)} = \ & e_1 - e^1 \\
v_{(1,0)}' = \ & e_1 - e^1 -e^2 \, .
\end{aligned}
\end{equation}
The remaining $11$ pairs of points $\hat{p}_i$ and $\hat{p}_i'$ are grouped around the $11$ points $\Delta_{het}=0$. The M-theory dual of a geometric compactification of heterotic string theory (i.e., a compactification not involving patching by T-dualities) corresponds to bringing the $\hat{p}_i$ and $\hat{p}_i'$ close together ($\lambda \rightarrow \infty$), so that we can effectively ignore the monodromies encountered when passing in between two such points, i.e., the `T-duality' monodromies swapping $\tau$ and $\sigma$. 

As we have seen, the monodromy group generated by paths encircling the loci $\Delta_{het}=0$ is the group of transformations $M(a,b,c,d)$ \eqref{eq:matrixofdoom} which are equal to the product of Picard-Lefschetz transformation associated with the cycles \eqref{eq:my_bicycles_have_vanished}. By transforming the cycles \eqref{eq:my_bicycles_have_vanished} with a general transformation \eqref{eq:matrixofdoom}, we can find the vanishing cycles in $H_2(S_-,\mathbb{Z})$ corresponding to the $(p,q)$ vanishing cycle on the heterotic side
\begin{equation}
\begin{aligned}
v_{(p,q)} = &(2 p q+1 ) e_1  + (2 p q-1)e^1 -2 p^2 e_2  + 2 q^2e^2 \\
v_{(p,q)}' = & (p^3 q+p^2+2 p q+1)e_1  + (p^3 q-p^2+2 p q-1) e^1 -p^2 (p^2+2)e_2, + ((p^2+2) q^2-1) e^2\, .
\end{aligned}
\end{equation}
Equivalently, these are the orbits of \eqref{eq:my_bicycles_have_vanished} under the monodromy group $\{M(a,b,c,d)\}$. This pair of vanishing cycles generates a monodromy map (via Picard-Lefschetz) which is the equal to \eqref{eq:matrixofdoom} (after rewriting $a,b,c,d$ in terms of $(p,q)$) and the appropriate conjugate of $T_\tau T_\sigma^{-1}$, \eqref{eq:monoDelta_E}. 

Using the picture of string junctions, we have seen how an $E_8$ worth of open discs was constructed for an open $dP_9$ in the section \eqref{sec:StringJunk}. These discs all correspond to classes in the second homology of $dP_9$ relative to the $(1,0)$ cycle over a base point, the fiber over which is removed to obtain the open $dP_9$. Fixing a reference point on the base of $Z_-$ (again, this is naturally chosen as the point   over which the fiber is excised when forming $X_- = Z_- \setminus S_-^0$), the completed structure of such cycles is lifted to two sets of classes in the relative homology of $H_3(Z_-,S_-)$, each of which is isomorphic to $E_8$ (but recall that this map is no group homomorphism). These two sets contain cycles that restrict to, respectively, $e_1 - e^1$ and $e_1 - e^1-e^2$ on $S_-^0$. In the same way as the open discs on the heterotic side are obtained by capping off an $\mathbb{S}^1$ at one end of an half-open interval,
each such cycle will be represented by (the closure of) a submanifold isomorphic to $\mathbb{R}^3$, formed by capping off an $S^2$ at one end of a half-open interval. 

In lifting the structure of cycles present on the heterotic side, we have  ignored the monodromies associated with the points at $\delta\delta'=0$. This is in so far justified, as we may think of working in region of moduli space where these points are separated from the set of points $\Delta_{het}=0$ we employed in our construction. However, as the monodromies $M_\sigma$ act non-trivially on the cycles $v_{(p,q)}$ and $v_{(p,q)}'$, the location of the points at $\delta\delta'=0$ will in general interfere with our construction. We will come back to the physical relevance of this feature below.

\subsection{The Geometry of $S_+$ and $Z_+$}

In contrast to {$S_-$}, the two $E_8$ summands are contained in the transcendental lattice of $S_+$ instead of the Picard lattice. This means that the periods of the cycles in $S_+$ spanning the two $E_8$ summands vary over the base of $Z_+$, which encodes the twisting of an $E_8\times E_8$ vector bundle $\hbun$ on the heterotic side. The threefold $Z_+$ is given as an algebraic threefold by 
\be \label{eq:weierstrassZ+}
y^2 = x^3 + f_{4,8}(\hat{z},z) x + g_{6,12}(\hat{z},z) \; ,
\ee
where now $f_{4,8}$ and $g_{6,12}$ are generic polynomials of the indicated degree. One can check that the K3 fiber $S_+$ degenerates with an $A_1$ singularity over $264$ points in the $\mathbb{P}^1$ base of $Z_+$. Taking the appropriate version of the degeneration limit $\lambda\rightarrow \infty$ as in the last section, $24$ of these $264$ points pairwise coincide to realize the same algebraic structure as found for $Z_-$ and $S_-$, whereas the remaining $240$ points encode the bundle data. Again, we may think of separating these $240$ from the $12$ pairs of points $\hat{p}_i,\hat{p}_i'$ which pairwise come together to encode the geometry of the heterotic dual. Furthermore, note that $Z_-$ can be obtained as a particular singular limit of $Z_+$ in which $f_{4,8}$ and $g_{6,12}$ are tuned appropriately. In such a tuning, the $240$ monodromy points associated with the bundle $\hbun$ merge  in groups of $20$ at $12$ points associated with $\delta\delta'=0$\footnote{ This is a degeneration of $\hbun$ to small instantons on the heterotic side.}.

These observations have two important consequences. The first consequence is that we can repeat the same analysis done for $Z_-$ if we ignore the monodromies associated with the bundle data. As $T_+$ contains $U_1 \oplus U_3$ instead of $U_1\oplus U_2$, we associate the $(1,0)$ cycle in the $T^2$ fiber of the geometry of the dual heterotic compactification with 
\begin{equation}\label{eq:my_bicycles_have_vanished_again} 
\begin{aligned}
v_{(1,0)} = & e_1 - e^1 \\
v_{(1,0)}' = & e_1 - e^1 -e^3 \, 
\end{aligned}
\end{equation}
in $S_+$. As before, we find two copies of $E_8$ represented by relative homology cycles in $H_3(Z_+,S_+)$, one copy restricts to $e_1 - e^1$ on $S_+^0$ and the other restricts to $e_1 - e^1-e^3$ on $S_+^0$.

The second consequence is that the monodromies associated with the $240$ points on the base of the K3-fibration of $Z_+$ related to the bundle $\hbun$ act non-trivially on $v_{(p,q)}$ and $v_{(p,q)}'$. Hence the properties of the cycles we have constructed will depend on the location of those points.

\subsection{The Associative Submanifolds}
\label{sec:g2lift}

The discussion of the geometries of $Z_\pm$ from the point of view of the fibrations of $S_\pm$ now allows us to lift the set of sections $\sigma_{\hat \gamma \gamma}$ directly to the $G_2$-manifold $J$. As we have seen in section \ref{sec:StringJunk}, each cycle $\sigma_{\hat \gamma \gamma}$ is realized in the SYZ picture of the Schoen Calabi-Yau threefold $\hets$ by glueing two discs sitting in $H_2(V_\pm,T^2_\pm)$. These discs were in turn constructed by fibering an $\mathbb{S}^1$ over an appropriate tree-like graph in the base of $V_\pm$. Crucially, these asymptotic $\mathbb{S}^1$s are required to match up when glueing $V_+$ to $V_-$ to get back the Schoen. This leads to identifying these (uniquely) with the $\mathbb{S}^1$ that has a non-trivial fibration on both $V_\pm$, i.e., the one with coordinate {$u_5$} in figure \ref{fig:good_fibrations}. 

A similar structure is in place for $J$. As we have discussed above, there exists a subset $\{\Sigma_{\gamma}^+\}$ of $H_3(Z_+,S_+^0)\cong E_8$ as well as $\{\Sigma_{\hat{\gamma}}^-\}$ of $H_3(Z_-,S_-^0)\cong E_8$, both of which restrict to $e_1-e^1$ on $S_+^0 \cong S_-^0$. In the same way as the open discs of $\hets$ are glued together to a cycle in $\hets$ we can glue the relative homology cycles $\{\Sigma_{\gamma}^+\}$ and $\{\Sigma_{\hat\gamma}^-\}$ on $X_\pm$ to cycles $\Sigma_{\hat\gamma \gamma}$ of $J$. By duality, it follows that the corresponding classes contain a unique associative representative.\footnote{ Indeed, as we shall see below the $\mathbb S^1$ with coordinate $u_5$ is dual to an $\mathbb S^2$ calibrated by $\text{Im} \Omega_{S_{\pm}}$. By duality, the latter is fibered over the same tree-like graphs thus giving rise to special lagrangian thimbles in $X_\pm$, which glue to associatives of J because the $G_2$ structures $\Phi_{3,\pm}$ on $\mathbb S^1 \times X_\pm$ are respected by the TCS glueing morphism.}

Recall that the realization of the curves $\sigma_{\gamma, \hat{\gamma}}$ in terms of string junctions is merely a reconstruction of the sections of the double elliptic fibration on $\hets$. This implies that there is unique representative among the string junctions in the same homology class which reproduces the holomorphic section and that this representative has the topology of a two-sphere. In the representation in terms of string junctions, these cycles are realized as an $\mathbb{S}^1$ sitting over tree-like graphs (which may be a simple interval) collapsing to a point at each end of the graph and nowhere else on it. In the same fashion, the three-cycles $\Sigma_{\hat\gamma \gamma}$ are realized as two-spheres sitting over the same tree-like graphs collapsing to a point at each end of the graph and nowhere else. Hence we expect to have a unique associative representative by mapping the string junction reproducing $\sigma_{\gamma, \hat{\gamma}}$ under the duality in this way. Such associatives furthermore have the topology of rational homology three-spheres, in nice agreement with the result of \cite{Harvey:1999as}. 
We therefore conjecture:
\\
\\
\emph{For every element $(\gamma,\hat\gamma)$ of $E_8 \oplus  E_8$ there is a pair of three-chains   $\Sigma_{\gamma}^+$ and $\Sigma_{\hat \gamma}^-$ on $Z_\pm$ with boundaries $e_1-e^1$ (the unique effective $-2$ curve in $T_+\cap T_-$) in $S_\pm^0$, which can be glued to a three-cycle $\Sigma_{\gamma \hat\gamma}$ in $H_3(J)$. We conjecture that the class of this three-cycle contains a unique associative representative that has the topology of a three-sphere.} 
\\
\\
As we have seen, there is another subset of relative homology cycles isomorphic to $E_8$ for both $Z_+$ and $Z_-$. These are such that they restrict to $e_1 - e^1-e^2$ on $S_+^0$ and $e_1-e^1-e^3$ on $S_-^0$. In contrast to the associatives we have constructed, these cannot be glued as they do not match on the overlap $S_- \cap S_+$. We should emphasize at this point, that the part of the conjecture related to the calibration is entirely inferred from the duality chain.

There is an alternative presentation of the associatives in the above conjecture, which has a beautiful relation to the way the corresponding contributions to the superpotential appear on the heterotic and F-Theory sides. What makes the geometries of $\hets$ and $\Fth$ so special, is that both of them are fibered by a calibrated (holomorphic) $T^4$ over a calibrated base, which is $\mathbb{P}^1$ for $\hets$ and $\mathbb{P}^1\times \mathbb{P}^1$ for $\Fth$. {Consequently, the superpotential contributions we are interested in are given by holomorphic sections.}  Note in particular that in the case of F-Theory, the sections of the holomorphic $T^4 = \mathbb{E}\times \hat{\mathbb{E}}$ fibration is such that $\mathbb{E}$ has a unique section, while there is an $E_8$ worth of sections of the elliptic {$\hat{\mathbb{E}}$} fibration. Furthermore, while the M5-branes in M-Theory picture of F-Theory are wrapped on sections of the elliptic fibration with fiber $\hat{\mathbb E}$ only, if we choose the picture of Euclidean D3-brane instantons to describe the generation of the superpotential, these branes are wrapped on holomorphic sections of the fibration with fiber $T^4$. 

This begs the question if a similar structure is in place for $J$ and we wish to answer this question in the affirmative. In the study of mirror symmetry for TCS $G_2$-manifolds, \cite{Braun:2017ryx,Braun:2017csz} conjectured the existence of a coassociative $T^4$-fibration which plays the role of the SYZ-fiber in this context. The fibers of this fibration can be seen in the Kovalev limit of a TCS $G_2$-manifold as glued from the SYZ-fibers of the two Acyl Calabi-Yau manifolds $X_\pm$ times the auxiliary circles $\mathbb{S}^1_{e\pm}$. In the neck region of $J$, these $T^4$-fibers become the SYZ-fibers of the asymptotic K3 surface $S_{\pm}^0$ times $\mathbb{S}^1_{e +} \times \mathbb{S}^1_{e -} =  \mathbb{S}^1_{b +} \times \mathbb{S}^1_{b -}$. Due to the Donaldson matching,  $\mbox{Im}\,\Omega^{(2,0)}_- = - \mbox{Im}\,\Omega^{(2,0)}_+$, the vectors $\mbox{Im}\,\Omega^{(2,0)}_+$ are purely contained in $U_1 \otimes \mathbb{R}$. This turns $e^1$ into the fiber of a special Lagrangian fibration and $e_1 - e^1$ into its section and this fibration is identified with the SYZ-fibration of the asymptotic K3 fibers. In the neck region, the $T^4$-fibers are hence described as 
\begin{equation}
\mathcal{F} =  e^1 \times \mathbb{S}^1_{b +} \times \mathbb{S}^1_{b -} \, .
\end{equation}
In contrast, the associative cycles we have constructed above are of the form $[$interval $\times\,\, (e_1 - e^1)\,]$, which means they are geometrically located in the perpendicular directions and intersect the fiber $\mathcal{F}$ in a unique point. We are hence led to conjecture:
\\
\\
\emph{The TCS $G_2$-manifold $J$ is fibered by a coassociative $T^4$ over a base with the topology of a rational homology three-sphere, and the $T^4$-fiber restricts to the SYZ-fiber of the Acyl Calabi-Yau threefold $X_\pm \times \mathbb{S}^1_{e \pm}$. This fibration has infinitely many associative sections $\Sigma_{\gamma \hat\gamma}$ which are isomorphic to the lattice $E_8\oplus E_8$. Furthermore, there is a group acting by translations on the coassociative $T^4$-fiber, which allows to add sections. Using this group law, the above isomorphism between sections and the lattice $E_8\oplus E_8$ becomes a homomorphism of abelian groups.}
\\
\\
Note that the base of the fibration being a three-sphere implies that the associative three-cycles $\Sigma_{\gamma \hat\gamma}$ must also have the topology of a three-sphere. 

Using the presentation \eqref{CohomologiesTCS} of the cohomology of a TCS $G_2$-manifold, let us now identify the classes of the homology three-cycles (or equivalently four-cycles in cohomology) in which we expect these associatives to be contained. To do this, recall the parametrization \eqref{eq:hetcurvesrigid} of the rigid curves on the heterotic side:
\begin{equation}\label{eq:sdotsp}
\sigma_{\gamma,\hat\gamma} = \sigma_\gamma\cdot \hat \sigma_{\hat\gamma} = 
(\gamma + \sigma_0 + n F) \cdot (\hat\gamma + \hat\sigma_0 + \hat{n} {F}) \,,
\end{equation}
which can be written using the intersection form of divisors on $\hets$ as
\begin{equation}\label{eq:sdotsprecast}
\sigma_{\gamma,\hat\gamma}= \sigma_0 \cdot \hat\gamma + \hat\sigma_0 \cdot \gamma + \sigma_0 \cdot \hat\sigma_0 + (\hat{n} \sigma_0  + n \hat\sigma_0 )\cdot F    \,.
\end{equation}
We can now infer how each of these terms is lifted when we go to the corresponding $G_2$-manifold. Before this, let us recall {from section \ref{sec:FHGsub}} that $H_3(M)$ is Poincar\'e dual to 
\begin{equation}\label{eq:h4details}
\begin{aligned}
H^4(J) & =H^3(Z_+) \oplus H^3(Z_-) \oplus (T_+ \cap T_-) \oplus \Lambda/(N_- + T_+ ) \oplus \Lambda/(N_+ + T_- ) \\
&  \qquad \oplus K_- \oplus K_+ \oplus H^4(S) \,,
\end{aligned}
\end{equation}
and see which terms contribute. First of all, we expect a contribution irrespective of the how we distribute $c_2(\hets)$ among the bundles and NS5-branes on the heterotic side. This rules out a contribution from $K_\pm$ as their existence depends on this distribution \cite{Braun:2017uku}. Furthermore, the contribution from $H^4(S)$ is Poincar\'e dual to the base of the $K3$-fibration of $M$, which becomes the base of the SYZ-fibration under the duality. We are hence restricted to terms from the first row of Equation (\ref{eq:h4details}).

The contributions $H^3(Z_\pm)$ come from cycles that are localized on the building blocks. They are the $G_2$ analogs to the usual string junctions on $dP_9$ realizing the $E_8$ root lattice and hence we wish to associate them with the corresponding terms $\sigma_0 \cdot \hat\gamma + \hat\sigma_0 \cdot \gamma$ on the heterotic side. The $E_8$ lattice is in particular generated by its roots $\alpha_i$, and we associate the corresponding three-cycles inside $H_3(X_-)$ ($H_3(X_+)$) by $\alpha_i$ ($\hat\alpha_i$).
The remaining terms are
\begin{equation}
\begin{aligned}
N_+& = U_2  &\hspace{1cm}N_-& = U_3 \oplus E_8 \oplus E_8 \\
T_+& = U_1 \oplus U_3 \oplus E_8 \oplus E_8 &\hspace{1cm} T_-& = U_1 \oplus U_2 \,,
\end{aligned} 
\end{equation}
so that 
\begin{equation}
\begin{aligned}
 T_+ \cap T_- &= U_1 \\ 
 \Lambda/(N_- + T_+ ) &= U_2 \\
 \Lambda/(N_+ + T_- )&= U_3 \oplus E_8 \oplus E_8\,.
\end{aligned}
\end{equation}
Let us label the generators of the three $U$-lattices as before by $e_i,e^i$ with $e_i \cdot e^j = \delta_i^j$. 

Any section of the double elliptic fibration of the Schoen manifold corresponds to an $\mathbb{S}^1$ in the SYZ-fiber capping off at the ends of an interval in the base. As argued already above, it must lift to a $\mathbb{P}^1$ in the K3-fiber with a similar behavior for the TCS $G_2$-manifold $J$. The unique choice for such a cycle with non-trivial monodromies on both sides is 
\begin{equation}
e_1 - e^1 \in  T_+ \cap T_-  \, .
\end{equation}
As we have seen, this cycle can cap off on both $X_-$ and $X_+$ in an $E_8$ worth of ways, corresponding to the multitude of sections $\sigma_{\hat\gamma}$ and $\sigma_\gamma$. Picking two specific zero sections $\hat\sigma_0$ and $\sigma_0$ corresponds to fixing a pair of relative three-cycles on $X_\pm$ restricting to $e_1-e^1$ on $S_\pm^0$. We denote this three-cycle of $J$ by $\Sigma_{\hat{0}0}$ and associate 
\begin{equation}
\hat\sigma_0\cap\sigma_0 \leftrightarrow  \Sigma_{0\hat{0}}\,, \qquad   \Sigma_{0\hat{0}}|_{S^0_\pm} = e_1 -e^1\, .
\end{equation}
This three-cycle sits in $H^4(J)$ via the term $T_+\cap T_-$ in \eqref{eq:h4details}. 

Finally, we need to discuss the lift of the remaining cycles in \eqref{eq:sdotsp}. In the SYZ-fibration on the heterotic side, they are characterized by fibering an $\mathbb{S}^1$ in the SYZ-fiber, with a non-trivial behavior only on one of the two sides, over an interval. We would hence like to associate such cycles with elements in $U_2$ or $U_3$, which in turn sit in the two remaining terms in the first row of \eqref{eq:h4details}\footnote{ In this particular example, this still leaves room for the appearance of the summand $E_8 \oplus E_8$ in $ \Lambda/(N_+ + T_- )$. We wish to argue that the correct identification is $U_3$. First observe that the geometric situation is completely symmetric between the two halves of $\hets$ and the circles ${u_1}$ and ${u_3}$, see figure  \ref{fig:good_fibrations}. This urges us to realize the same symmetry for $J$. Second, we will shortly present a generalization of our result to cases in which $\Lambda/(N_+ + T_- )$ contains $U_3\oplus G$ for an arbitrary sublattice $G$ of $E_8 \oplus E_8$. }. Secondly, there is nothing which distinguishes the $\mathbb{S}^1$ with coordinate ${u_3}$ from that with coordinate ${u_5}$ if we consider $X_-$ alone, and the same holds true for ${u_1}$ and ${u_5}$ with respect to $X_+$, see figure  \ref{fig:good_fibrations}. As we have already concluded that the $\mathbb{S}^1$ with coordinate ${u_5}$ is lifted to the effective $-2$ curve $e_1-e^1$ in $H^2(S^0)$, we are led to choose $e_2 - e^2$ and $e_3 - e^3$ as the restrictions of the three-cycles which comprise the lifts of $\hat\sigma_0 \cdot F$ and $\sigma_0 \cdot F$. Let us denote the associated three-cycles by $\Sigma_{0F}$ and $\Sigma_{F\hat{0}}$ . Note that they suffer from the same ambiguity as $\Sigma_{0\hat{0}}$ in that e.g.\ we need to choose a path in the base of $X_-$ to define how $e_3-e^3$ caps off to define $\Sigma_{0F}$. This is not unexpected, as these cycles are the lifts of $\sigma_0 \cdot F$, where the same ambiguity of choosing a zero-section is present. 
In particular, we are going to associate
\begin{equation}\label{eq:cycleidentification}
\begin{aligned}
\sigma_0 \cdot F &\leftrightarrow & \Sigma_{0F} &\qquad \qquad   \Sigma_{0F}|_{S^0_\pm} = e_2 -e^2\\
\hat\sigma_0  \cdot F & \leftrightarrow &  \Sigma_{F\hat{0}} &\qquad \qquad   \Sigma_{F\hat{0}}|_{S^0_\pm} = e_3 -e^3 \,.
\end{aligned}
\end{equation}
Note that all of the terms we have identified are associated with cycles contained in the groups $H^3(X_+) \oplus H^3(X_-)$, which is precisely how the first row of \eqref{eq:h4details} comes about in a computation of $H^4(J)$ in the Mayer-Vietoris sequence. In particular, the relevant contributions are \cite{Corti:2012kd}
\begin{equation}
T_+\cap T_- \oplus H^3(Z_+) \oplus H^3(Z_-) = \ker \left[(\beta^3_-\oplus \beta^3_+): H^3(X_+) \oplus H^3(X_-) \rightarrow H^2(S^0) \right] \,,
\end{equation}
where $S^0\cong S^0_+\cong S^0_-$, together with 
\begin{equation}
\begin{aligned}
&\Gamma/(N_- + T_+ ) \oplus \Gamma/(N_+ + T_- )  =  \\
&\mbox{coker} \left[ \left(\begin{array}{cccc}
       \beta^3_+ & 0 & 0 & \rho_- \\
        0 & \beta^3_- & \rho_+  & 0
      \end{array}\right) : H^3(X_+) \oplus H^3(X_-) \oplus H^2(X_+) \oplus H^2(X_-) \rightarrow H^2(S^0)\oplus H^2(S^0) \right]\, .
\end{aligned}
\end{equation}
This appearance and the associated identifications are a reflection of the fact that $\Sigma_{0\hat{0}}$ represents an arbitrary choice among all of the cycles $\Sigma_{\gamma\hat\gamma}$, as do $\Sigma_{0F}$ among $\Sigma_{\gamma F}$ and $\Sigma_{F\hat{0}}$ among $\Sigma_{F \hat\gamma}$.

In summary, adapting the reasoning of section \ref{sec:DGW} to the $G_2$ setting, we are led to identify the classes of the associatives three-cycles $\Sigma_{\gamma\hat\gamma}$ as 
\begin{equation}\label{eq:theassociativesinabasis}
\Sigma_{\gamma\hat\gamma} = \sum_{i=1}^8 \left(m_i \alpha_i +  \hat{m}_i \hat{\alpha}_i \right) + \Sigma_{0\hat{0}} + \hat{n}\Sigma_{F\hat{0}}  + n \Sigma_{0F} 
\end{equation}
for a choice of three-cycles $\alpha_i,\hat\alpha_i$ in $H^3(Z_+)$ and $H^3(Z_-)$ and with 
\begin{equation}\label{eq:ndoesnotmatter2}
\begin{aligned}
n = \sum_i (m_i)^2 - (m_1 m_2 +\cdots + m_3 m_8) \\
\hat{n} = \sum_i (\hat{m}_i)^2 - (\hat{m}_1 \hat{m}_2 +\cdots + \hat{m}_3 \hat{m}_8) \,. 
\end{aligned}
\end{equation}
With this geometric identification of the three-cycles dual to the calibrated cycles that generate the instanton corrections in heterotic and F-theory, we are now in a position to study the associated corrections in the M-theory compactification on $J$.

\subsection{The Superpotential} 
\label{sec:superpot}

The superpotential corrections arising from M2-brane instantons for M-theory on $G_2$-manifolds has been discussed in \cite{Harvey:1999as}. The matching of our results under the various dualities gives a consistency check for the approach proposed there. The computation of the M-theory superpotential is based on the assumption that the contribution of an M2-brane wrapping an associative three-cycle can be approximated by coupling the M2-brane with the supergravity background corresponding to the $G_2$ holonomy geometry, and performing the path-integral on the phase space for the resulting three-dimensional theory. The result is that for each associative three-cycle $\Sigma$ which is a rigid rational homology three-sphere, the superpotential receives a contribution of
\begin{equation}
\Delta W \propto {\mathcal D}^{\text{sugra}}_\Sigma \, |H_1(\Sigma,\mathbb Z)| \text{ exp}\left( 2 \pi i\int_\Sigma C + i \Phi_3 \right) \,,
\end{equation}
where the proportionality factor is a universal constant coefficient (powers of 2 and $\pi$) that takes care of the overall normalization of the superpotential.  The prefactor ${\mathcal D}^{\text{sugra}}_\Sigma$ takes into account the contribution from 1-loop determinants that involve the fields that are part of the supergravity background. This term can be treated as a constant universal factor whenever it is legitimate to approximate the M2-brane as an elementary brane with no account of backreaction. As we shall see below, these effects are important for the continuity of the superpotential based on the transitions in the spectrum of associative three-cycles along the $G_2$-moduli space. It is important to remark that the superpotential can receive contributions from supersymmetric three-cycles that have $b_1(\Sigma)>0$ and that are non-rigid, that comes from higher order terms in the DBI action responsible for soaking up the extra zero-modes. Here we adopt an adiabatic approximation to the M2-brane dynamics and these higher oder terms can be neglected (see below for more on this point). It is also possible to have contributions from sectors with multiple wrappings.

Having identified the associative three-cycles in the last section, we can repeat the same computation as before to find the superpotential. In particular, if we use a basis of $H^3$ dual to the cycles $\{\alpha_i,\hat\alpha_i,\Sigma_{0\hat{0}},\Sigma_{F\hat{0}},\Sigma_{0F}\}$ to expand $C + i \Phi_3$ (we indicate dual elements by $*$)
\begin{equation}
C + i \Phi_3 = \sum_i \left( \alpha_i^*   \omega_i + \hat\alpha_i^{*}   \hat\omega_i \right)+ \Sigma_{F\hat{0}}^* \tau + \Sigma_{0F}^* \hat\tau + \Sigma_{0\hat{0}}^* z + \cdots
\end{equation}
we immediately find
\begin{equation}\label{eq:superduperMTheory}
\begin{aligned}
W & =  \sum_{\Sigma_{\gamma\hat\gamma}} G(\gamma\hat\gamma) \exp \left[2\pi i \int_{\Sigma_{\gamma\hat\gamma}} C + i \Phi_3 \right] \\
  &=  \sum_{m,\hat{m}\in \mathbb{Z}^8 \times \mathbb{Z}^8  } G(\gamma\hat\gamma) \exp 2\pi i \left[z + n \tau + \hat{n}\hat\tau +  \sum_i m_i \omega_i +  \hat{m}_i \hat\omega_i\right]   \,,
\end{aligned}
\end{equation}
where the dependence on $n$ and $\hat n$ is taken into account by \eqref{eq:ndoesnotmatter2}, and the prefactors $G(\gamma\hat\gamma)$ are given by 
\begin{equation}
G(\gamma\hat\gamma) \propto {\mathcal D}^{\text{sugra}}_{\Sigma_{\gamma\hat\gamma}} \qquad\qquad |H_1({\Sigma_{\gamma\hat\gamma}},\mathbb Z)| = 1
\end{equation}
for three-cycles $\Sigma_{\gamma\hat\gamma}$ that correspond to primitive vectors of the $\mathbb{Z}^8 \times \mathbb{Z}^8$ lattice, while for three-cycles $\Sigma_{\gamma\hat\gamma}$ that correspond to non-primitive vectors the prefactors are complicated by taking into account the effects due to multiple wrapping.\footnote{  Recall that a given element $\gamma = \sum n_i e_i$ of an integer lattice generated by the vectors $e_i$ is said to be primitive whenever $gcd(n_i) =1$. Of course these elements are not affected by ambiguities arising from multiple wrapping.}

Assuming that for special values of the moduli the prefactor $G(\gamma\hat\gamma)$ is universal, the expression \eqref{eq:superduperMTheory} evaluates to 
\begin{equation}
W=  e^{2 \pi i z}\, \Theta_{E_8}(\tau,{\omega})\,\, \Theta_{E_8}(\hat\tau,{\hat\omega}) 
\end{equation}
at that point in moduli space, which matches the expressions found in heterotic and F-theory under analogous assumptions with regards to the universality of the prefactor.

{Let us now discuss the prefactor $G(\gamma\hat\gamma)$ in more detail.} As we have argued above, the $12$ monodromy points located at $\delta\delta'=0$ in $X_-$, which encode the heterotic NS5-branes or D3-branes in F-theory, induce a monodromy action which acts non-trivially on the curve $e_1-e^1$ and its images under $M_\tau M^{-1}_\sigma$. The same happens for the $240$ monodromy points on $X_+$ associated with the data of the heterotic bundle $\hbun$. This means in particular that the vanishing cycles of the degenerations of the K3 fiber over these points in general intersect the cycle $e_1-e^1$ (and its images under the monodromy group) used to construct the associative submanifolds $\Sigma_{\gamma \hat\gamma}$. As we move in the moduli space of the $G_2$-manifold $J$ while staying in the Kovalev limit, the loci in the base $S^2\setminus$pt of $X_\pm$ at which the K3-fiber degenerates will move as well. In particular, such points $\hat{p}$ come close, or even coincide, with an associative $\Sigma_{\gamma \hat\gamma}$. This implies that the minimal volumes or even the existence of the associative cycles we have constructed will depend on the positions of these monodromy points. This interplay is not unexpected, as the prefactors of the contributions of the heterotic worldsheet instantons depend on both the positions of the NS5-branes, as well as the bundle moduli of $\hbun$ of $\hets$. In particular, it is known that the contribution of an {D3-brane} instanton in F-Theory is absent due to an extra zero mode if a D3-brane is moved on top of it \cite{Ganor:1996pe}. 

It is well-known that the spectrum of associative three-cycles depend on the position on the moduli space of a given $G_2$-manifold. For one-parameter families $J_t$, $t\in \mathbb R$, of $G_2$-manifolds, there are six possible behaviors that have been suggested by Joyce \cite{Joyce:2016fij}, schematically these are the following:
\begin{itemize}
\item[A.)] Canceling non-singular associatives with opposite signs: \\
This is the geometric analogue of a creation/annihilation: for $t< t_0$ there is no associative, at $t=t_0$ there is a single one $\Sigma_0$, at $t>t_0$ there are two $\Sigma^t_i$, $i=1,2$, such that $\text{lim}_{t\to t_0} \Sigma^t_i = \Sigma_0$ but have opposite orientations.
\item[B.)] Intersecting associatives give connected sum:\\
 This is the geometric analogue of the formation of a bound-state: for $t< t_0$ there are two unobstructed associatives $\Sigma^t_i$, $i=1,2$, that do not intersect, at $t=t_0$ these intersect at a point, at $t>t_0$ there is a third associative $\Sigma^t_3 = \Sigma^t_1 \# \Sigma^t_2$ such that $[\Sigma^t_3] =[\Sigma^t_1]+[\Sigma^t_2] $.
\item[C.)] Self-intersecting associative give connected sum $\Sigma \# (\mathbb{S}^1 \times S^2)$:\\
 This case and the following are similar, and correspond to the bubbling of an excited state: for $t< t_0$ there is a single unobstructed  associative $\Sigma^t$, at $t=t_0$ the associative $\Sigma^{t_0}$ has a point of transverse self-intersection, for $t> t_0$ there is an additional associative which is given by the connected sum of $\Sigma^t \# (\mathbb{S}^1 \times S^2)$.
\item[D.)] Self-intersecting associative give connected sum $\Sigma \# \Sigma$:\\
 For $t< t_0$ there is a single unobstructed  associative $\Sigma^t$, at $t=t_0$ the associative $\Sigma^{t_0}$ has a point of transverse self-intersection, for $t> t_0$ there is an additional associative which is given by the connected sum of $\Sigma^t \# \Sigma^t$.
\item[E.)] Three associatives $\Sigma^t_{1,2,3}$ form $\Sigma^{t_0}_0$ with conical singularity: \\
This process is the geometric analogue of a decay/a formation of a bound state.
\item[F.)] Multiple covers:\\
This is the geometric analogue of a brane recombination: there is a family of associatives that at a special point coincide with the multiple cover of another one.
\end{itemize}
Each of these geometric phenomena can affect the superpotential contributions. Notice that the requirement that the superpotential is a smooth function of the moduli entails that all these contributions have to be modulated by the prefactors. For instance, consider case B, and assume that all associatives involved are rigid rational homology three-spheres. Running time in reverse order: the contribution of the cycle $\Sigma_3$ disappears from the potential after $t_0$. This is compatible with the above mentioned continuity only if the corresponding backreaction prefactor renders this transition smooth.

Though the background and instanton zero modes of \cite{Donagi:1996yf} determined
that our M2-instantons wrap rigid associative submanifolds that are homology
three-spheres, a much richer set of possibilities are available in M-theory,
as required by various dualities. Given the importance of M2-instantons
in M-theory compactifications, we would like to briefly catalog some of
the possibilities. Via duality with examples in F-theory
where non-rigid {D3-instantons} contribute due to the flux-lifting of 
deformation zero modes \cite{Bianchi:2011qh}, there may also exist cases where M2-instantons 
on non-rigid associatives contribute. Fluxed instantons in F-theory also give
rise to a number of other effects  \cite{Heckman:2008es, Grimm:2011dj ,Marsano:2011nn,Kerstan:2012cy,Martucci:2015oaa}, such as lifting charged chiral zero modes,
that could arise in M-theory duals. Instantons in heterotic/F-theory duals 
exhibit dependence on vector bundle \cite{Buchbinder:2002pr,Buchbinder:2002ic,Buchbinder:2017azb}/ seven-brane moduli \cite{Cvetic:2011gp,Cvetic:2012ts} that must arise as
$G_2$-moduli dependence in M-theory duals; by duality, superpotential zeroes
of this moduli dependence should be associated with singularity enhancement
(perhaps pointlike) in the M2-instanton worldvolume. Similarly, zero modes
in heterotic/F-theory duals that are Ganor strings
introduce dependence on NS5-brane/D3-brane moduli \cite{Ganor:1996pe,Baumann:2006th} that should arise also in $G_2$-moduli, as we discuss
at length. In singular limits of
$G_2$-compactifications that exhibit charged chiral matter, duality with type
IIA should allow for those superfields to appear in gauge-invariant combinations
in instanton prefactors \cite{Blumenhagen:2006xt,Ibanez:2006da,Florea:2006si}, which could be of
phenomenological importance. The development of techniques that allow for
efficient study of zero modes at large numbers of $G_2$-moduli, analogous
to the type IIB techniques in \cite{Braun:2017nhi}, is of great interest for studying the
landscape of $G_2$ compactifications.
Each of these possibilities, as well as others, should exist in some form
in M-theory, in particular we conjecture these have counterparts within the framework of Joyce's conjectures \cite{Joyce:2016fij}
on the moduli dependence of associative submanifolds we briefly mentioned above.

\subsection{Generalizations}

The presentation we have given lends itself to a generalization of our result to all TCS $G_2$-manifolds {(with elliptically fibered building blocks)}, which we will briefly explain. First of all, we might consider a different bundle $\hbun = \oplus_i \hbun_i$  with structure group contained in $E_8\times E_8$, together with different values of $\ch_2(\hbun_i)$, i.e., different numbers of NS5-branes wrapped in $\mathbb{E}$ and $\hat{\mathbb{E}}$. Following \cite{Braun:2017uku}, all of these will be captured by TCS $G_2$-manifolds for which both $S_+$ and $S_-$ are elliptically fibered such that 
\begin{equation}
\begin{aligned}
N_+& = U_2 \oplus G_+^\perp &\hspace{1cm}N_-& = U_3 \oplus G_-^\perp \\
T_+& = U_1 \oplus U_3 \oplus G_+ &\hspace{1cm} T_-& = U_1 \oplus U_2 \oplus G_-
\end{aligned}
\end{equation}
with $G_\pm\supset E_8 \oplus E_8$ and $G_\pm^\perp$ the orthogonal complement of $G_\pm$ in $E_8\oplus E_8$. It follows that
\begin{equation}
\begin{aligned}
 T_+ \cap T_- &\supseteq U_1 \\ 
\Lambda /(N_- + T_+ ) &\supseteq U_2 \\
\Lambda/(N_+ + T_- ) &\supseteq U_3\,. \\
\end{aligned}
\end{equation}
Note that this generalized setup includes singular $G_2$-manifolds and does not need to have an (obvious) F-Theory dual, as $\hbun$ does not need to be flat on either $\mathbb{E}$ or $\hat{\mathbb{E}}$. However, the geometry on the heterotic side is still the same, so that we expect the contribution from world-sheet instantons on $\sigma_\gamma \cdot \hat\sigma_{\hat\gamma}$ to be still present. Correspondingly, all of the structure we used to find with respect to the associatives $\Sigma_{\gamma\hat\gamma}$ are still in place. As $S_\pm$ are elliptic, we can consider the degeneration limit $\lambda \rightarrow \infty$ and construct  
$\Sigma_{\gamma\hat\gamma}$ by considering $e_1 - e^1$ fibered over an interval because $T_+ \cap T_-$ always contains $U_1$. 

The interplay between the cycles coming pairwise together around $\Delta_{het}=0$ that were used to construct the $\Sigma_{\gamma\hat\gamma}$, and the remaining monodromy points of the K3-fibers $S_\pm$, is then expected to account for the different effects of the  configurations of bundles and NS5-branes on the contribution of the world-sheet instantons to the superpotential. Furthermore, the identification of the classes \eqref{eq:theassociativesinabasis} will remain completely unchanged. We are hence led to {the conclusions and} conjecture:
\\
\\
\emph{Let $J$ be a TCS $G_2$-manifold (possibly singular) which is glued from two Acyl Calabi-Yau manifolds $X_\pm$ which are fibered by elliptic K3 surfaces $S_\pm$. For every element $(\gamma,\hat\gamma)$ of $E_8 \oplus  E_8$, there is a pair of three-chains $\Sigma_{\hat\gamma}^-$ and  $\Sigma_{\gamma}^+$ on $Z_\pm$ with boundaries $e_1-e^1$ in $S_\pm^0$, which can be glued to a three-cycle $\Sigma_{\gamma \hat\gamma}$ in $H_3(J)$. We conjecture that the class of this three-cycle contains a unique associative representative that has the topology of a three-sphere.}
\\
\\
Furthermore, we may extend the conjecture about such associatives being sections of a coassociative $T^4$-fibration, which is present on any such $G_2$-manifold (at least in the Kovalev limit):
\\
\\
\emph{Let $J$ be a TCS $G_2$-manifold (possibly singular) which is glued from two Acyl Calabi-Yau manifolds $X_\pm$ which are fibered by elliptic K3 surfaces $S_\pm$. Then $J$ is fibered by a coassociative $T^4$ over a base with the topology of a rational homology three-sphere, and the $T^4$-fibers restrict to the SYZ-fibers of the Acyl Calabi-Yau threefolds $X_\pm \times \mathbb{S}^1_{e \pm}$. This fibration has infinitely many associative sections $\Sigma_{\gamma \hat\gamma}$ which are isomorphic to the root lattice of $E_8\oplus E_8$. Furthermore, there is a group acting by translations on the coassociative $T^4$-fiber, which allows to add sections. Using this group law, the above isomorphism between sections and the lattice $E_8\oplus E_8$ becomes a homomorphism of abelian groups.}
\\
\\
It is tempting to extend this even further by noting that in fact \emph{any} TCS $G_2$-manifold should have a coassociative $T^4$-fibration at least in the Kovalev limit. However, it is hard to see how to construct the cycles $\Sigma_{\gamma \hat\gamma}$ in this case. The reason is that while $T_+ \cap T_-$ must always contain a class of positive self-intersection, it does not necessarily contain a $-2$ curve. Hence a possible analog of $e_1-e^1$, the fibration of which gave us the cycles $\Sigma_{\gamma \hat \gamma}$, does not need to be present.

\section{Conclusions and Outlook}
\label{sec:outlook}

A large class of compact $G_2$ holonomy manifolds may be constructed as twisted connected sums of asymptotically cylindrical Calabi--Yau threefolds \cite{Corti:2012kd, MR3109862,MR2024648}; each such manifold naturally comes equipped with a {topological} K3-fibration. Further specializing to the case where the building blocks of these $G_2$-manifolds are elliptically fibered, fiberwise duality can be used  to map M-Theory on such TCS $G_2$ manifolds to the heterotic string theory compactified on the Schoen Calabi--Yau threefold, as well as an F-theory model associated to a K3-fibered Calabi--Yau fourfolds over $dP_9$ \cite{Braun:2017uku}. 

Our main objective in the present paper has been to use this duality map to construct certain non-perturbative effects in M-theory that arise from Euclidean M2-branes wrapping associative three-cycles in the $G_2$-manifold. Starting from the work of Donagi--Grassi--Witten \cite{Donagi:1996yf}, we have identified an infinite class of three-cycles on elliptically fibered TCS $G_2$ holonomy manifolds that we conjecture, based on the duality chain, to have associative submanifold representatives with the topology of rational homology spheres. We have also argued that our conjecture has an obvious extension to any TCS $G_2$-manifold glued from building blocks with elliptic fibrations, which are dual to heterotic strings on $\hets$ with different bundles.

This conjectured existence of infinitely many associative three-cycles has implications for both physics and mathematics. Non-perturbative effects are ubiquitous in compactifications of string theory, and may have dramatic consequences for the stability of a classical solution of the theory. However, in many situations, these effects are difficult to study, owing to the lack of a geometric description. In contrast, the non-perturbative effects we identify in  M-theory compactifications on $G_2$-manifolds are completely geometrical. We thus find an ideal setting, where non-perturbative effects in physics can be studied using geometric tools, despite the fact that only a minimal amount of supersymmetry is preserved.  From a mathematical perspective, it should be noted that there are many outstanding questions regarding the geometry of $G_2$ holonomy manifolds and their calibrated cycles. Our indirect reasoning, based on string dualities, has led to a conjecture about the existence of infinitely many associative three-cycles. There is obviously need for a rigorous mathematical analysis of this conjecture, and we hope that the present work may serve as an inspiration for such a study.

Our findings {are related to} Joyce's recent conjectures regarding the counting of associative three-cycles on $G_2$-manifolds  \cite{Joyce:2016fij}. The  M2-instantons on associative three-cycles that we construct  give rise to a superpotential, {of the type} previously  discussed by Acharya \cite{Acharya:1998pm,Acharya:2000gb} and Harvey--Moore \cite{Harvey:1999as}. In the present paper, we used the M-theory/heterotic/F-theory duality chain to argue that there are an infinite number of classes that have a single associative submanifold representative and hence an infinite number of contributions to the superpotential. Based on the duality chain, we do not expect all such contributions to persist throughout the $G_2$ moduli space. This interesting dependence on the $G_2$ moduli that we have inferred from string theory matches Joyce's observation  \cite{Joyce:2016fij} that associative three-cycles in $G_2$-manifolds manifest six different type of wall-crossing behavior as the $G_2$ geometry is deformed, as summarized in section \ref{sec:superpot}. A related observation is applicable also to M5-branes, where the reduction results in a 3d $\mathcal{N}=1$ theory, $T[M_3]_{\mathcal{N}=1}$,  whose spectrum depends on the associative cycle $M_3$ \cite{TBASSN}.
It would be very interesting to use the duality chain in order to explore these different wall-crossing phenomena  in more detail.  A particularly intriguing proposal of Joyce is that it should be possible to define a moduli-invariant counting of associative $\mathbb{Q}$-homology three-spheres. We hope to test this proposal, using the dual perspectives of string theory, in the future.


\subsection*{Acknowledgments}

We thank Bobby Acharya, Chris Beasley, Julius Eckhard, Dominic Joyce, Cody Long, Luca Martucci, Julius Shaneson, Ben Sung, Eirik Svanes, Roberto Valandro, Timo Weigand and Jenny Wong for discussions. We {also} thank the Aspen Center for Physics for hospitality during the initial stages of the project. 
The  Aspen  Center  for  Physics  is  supported  by  National Science Foundation grant PHY-1607611, {and our visit there was enabled {in part} by}
Simons Foundation  grant \# 488629, which also supported the work of DRM reported on here. APB and SSN are supported by the ERC Consolidator Grant 682608 ``Higgs bundles: Supersymmetric Gauge Theories and Geometry (HIGGSBNDL)''. JH is supported by NSF grant PHY-1620526. The research of ML is financed by the Swedish Research Council (VR) under grant number 2016-03873. 


\appendix

\section{String Junctions for $dP_9$}
\label{app:StringJ}

In this section we explicitly compute the junctions in the heterotic picture of section \ref{sec:StringJunk} in a fixed $dP_9$ geometry\footnote{J.H. thanks A. Grassi and J.L. Shaneson for early discussions of string junctions 
in this geometry in $2012$.}, using small deformations of a Weierstrass model as systematically developed in \cite{Grassi:2013kha, Grassi:2014sda, Grassi:2014ffa}; we refer the reader to \cite{Grassi:2014sda} for explicit examples that are similar to this one.
Consider a Weierstrass model for $dP_9$, defined by 
\begin{equation}
f=(z+1)^4 \,,\qquad g = (z+1)^5 \,,
\end{equation}
with associated projection
\begin{equation}
dP_9 \xrightarrow{\pi} \mathbb{P}^1.
\end{equation}
In the Weierstrass model we have gone to a patch
of the $\mathbb{P}^1$ base.
Singular fibers exist over the discriminant locus $\Delta=4f^3+27g^2=0$,
with fiber types and locations given by
\be\ba
II^* \quad \hbox{at}\quad  &z=-1 \cr 
I_1\quad \hbox{at}\quad  &z=\frac{-2\pm 3\sqrt{3}i}{2}  \,.
\ea
\ee
Let us refer to these points as $\points_{E_8}$ and $\points_\pm$,
respectively. We will define the point $p$ to be $z=0$, which will
play a distinguished role in our study of relative homology.

\begin{figure}
\begin{center}
\centerline{
\includegraphics[scale=.8]{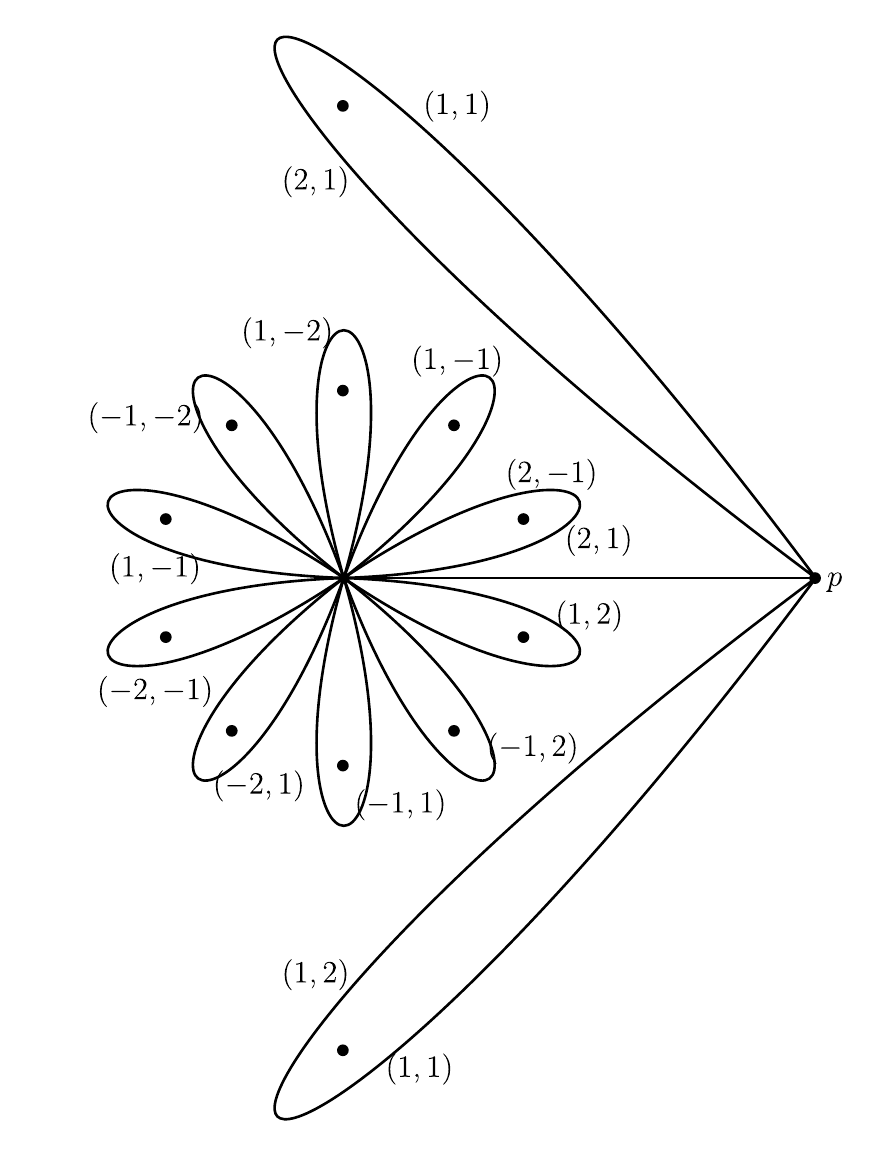}
\includegraphics[scale=.85]{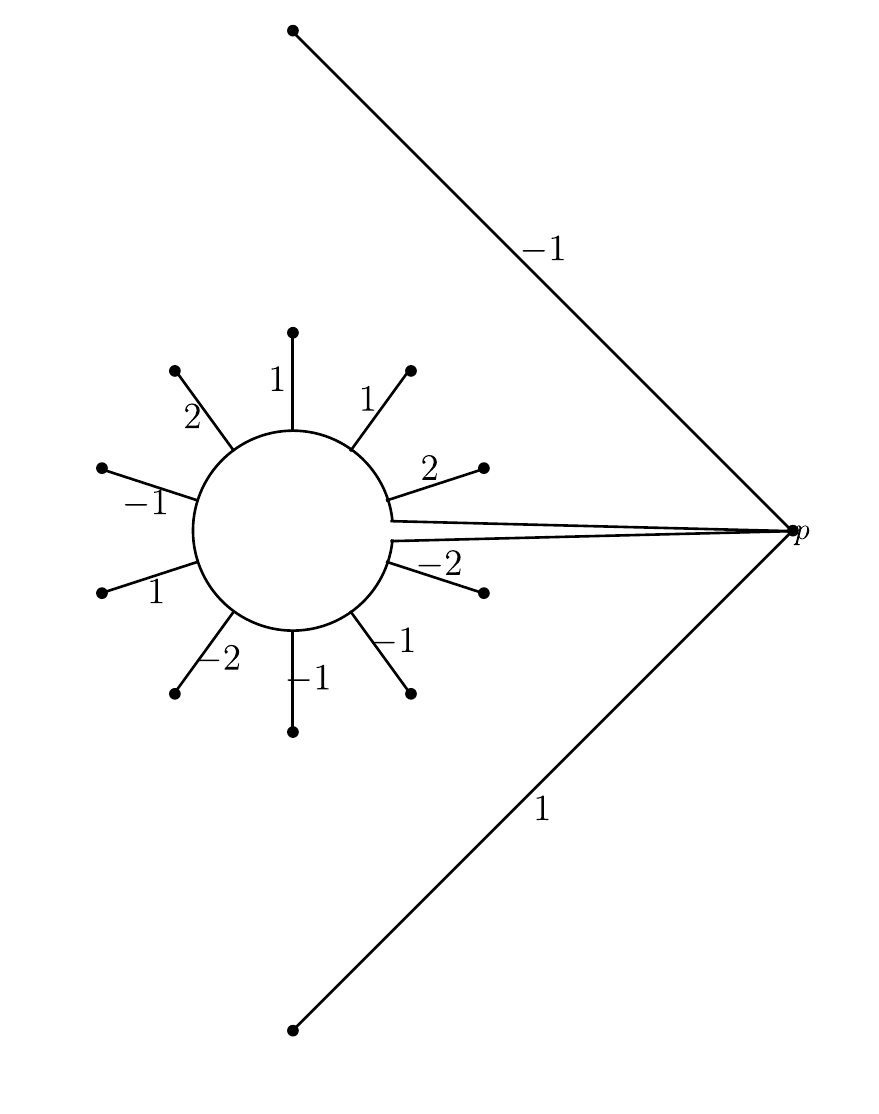}}
\end{center}
\caption{\emph{Left:} 
The torus $E$,
as represented by a loop around all
twelve defects. \emph{Right:}
$E$ represented as a junction, after
performing Hanany-Witten moves for
each loop in the figure at left. }
\label{fig:Ehwmove}
\end{figure}

Under perturbation $f\mapsto f + \epsilon$
the discriminant becomes a polynomial with $12$
distinct roots. If $\epsilon$ is sufficiently
small, the $II^*$ fiber splits into $10$ $I_1$
fibers collected around $\points_{E_8}$ and the other
$I_1$ fibers are perturbed slightly away from
$\points_\pm$. Explicitly doing so gives a picture of seven-branes as portrayed in
figure \ref{fig:Ehwmove},
where the points that form a circle
are the locations
of the $I_1$ fibers of the deformed $E_8$
and the points at top and bottom are the perturbations
of $\points_\pm$.
Henceforth, let us refer to $\points_\pm$ as the
locations of these perturbed $I_1$ fibers.

We now determine the vanishing cycles of the
$I_1$ loci. Let $E_p:=\pi^{-1}(p)$ be the smooth
elliptic fiber above $p$. We would like to study
two-cycles relative this elliptic fiber,
\begin{equation}
H_2(dP_9,E_p).
\end{equation}
Upon taking any path from $p$ to any of the $I_1$ loci
a one-cycle vanishes. The vanishing cycle depends on
the path, but we will compute the vanishing cycles
associated with particularly simple paths. Let
$\points_i$ with $i=1,\dots,10$ be the $I_1$ loci arising
from the deformation of the type $II^*$ fiber, beginning
with the upper right-most one and moving counter-clockwise. The
paths that we take to these defects are the straight
line from $p$ to $\points_{E_8}$, and then from $\points_{E_8}$
to $p\points_i$; recall that the fiber above $\points_{E_8}$ is 
smooth after the deformation. For the 
$I_1$ loci at $\points_\pm$, we simply take straight line
paths from $p$ to $\points_\pm$.

To determine the vanishing cycles, a basis of 
one-cycles must be chosen on $E_p$. In
the $x$-plane, the roots
of the Weierstrass cubic evaluated at $p$ appear as a triangle of points, one on the
left and two on the right,
where the roots are ramification points of
the double-cover $y^2=x^3+f x + g$. The torus
$E_p$ is a double cover of $\mathbb{C}$ with
four ramification points, which are the three
marked points and a point at infinity. Let $A$
be the cycle associated with the interval between
the left point and the bottom-right point, and $B$ the
cycle associated with the interval between the 
left point and the upper-right point\footnote{ Technically this only defines the cycles up to a sign, but the sign is irrelevant for the Picard-Lefschetz monodromy and the 
string junction analysis in the $dP_9$. We therefore
ignore this subtlety here.}. Give the paths that we chose
to the defects, a natural ordering of points is
\begin{equation}
\{\points_1,\points_2,\points_3,\points_4,\points_5,\points_6,\points_7,\points_8,\points_9,\points_{10},\points_-,\points_+\} \,.
\end{equation}
By direct computation, the associated ordered set of
vanishing cycles is
\begin{equation}
Z:=\{\gamma_1,\gamma_2,\gamma_3,\gamma_4,\gamma_5,\gamma_6,\gamma_7,\gamma_8,\gamma_9,\gamma_{10},\gamma_- ,\gamma_+\}=\{A,B,A,B,A,B,A,B,A,B,A,B\} \,,
\end{equation}
where $\gamma_i$ and $\gamma_\pm$ are the vanishing
cycles associated with $\points_i$ and $\points_\pm$.

As each path is followed to each $I_1$ locus,
the vanishing cycles form the cigar or Lefschetz
thimble in the geometry that we call $\Gamma_i$
and $\Gamma_\pm$, where
\begin{equation}
\Gamma_i, \Gamma_\pm \in H_2(dP_9,E_p) \,.
\end{equation}
To make notation easier, we define $\Gamma_{11}=\Gamma_-$
and $\Gamma_{12}=\Gamma_+$ and therefore
we may write some of the elements of $H_2(dP_9,E_p)$
as 
\begin{equation}
J = \sum_{i=1}^{12} J_i\, \Gamma_i \,.
\end{equation}
In the context of F-theory, these objects are known as
``string junctions", but geometrically they are just sums of
cigars, and thus are two-chains in this relative homology group. 
The boundary of such a $J$ is
\begin{equation}
a(J):=\partial J  \in H_1(E_p,\mathbb{Z})\,,
\end{equation}
which is referred to as the ``asymptotic charge''
in the physics literature.
In this case, representing the $A$-cycle by $(1,0)^T$ and the $B$-cycle
by $(0,1)^T$, the asymptotic charge is
\begin{equation}
a(J)=\begin{pmatrix} J_1+J_3+J_5+J_7+J_9+J_{11} \\J_2+J_4+J_6+J_8+J_{10}+J_{12} \end{pmatrix}\in H_1(E_p,\mathbb{Z}) \,.
\end{equation}
If $J$ has $a(J)=0$, i.e., it is a closed class, 
then $J\in H_2(dP_9,\mathbb{Z})$.
There is also a pairing
\begin{equation}
(\cdot,\cdot):  H_2(dP_9,E_p)\times H_2(dP_9,E_p) \to \mathbb{Z}
\end{equation}
that becomes the topological intersection product
on closed classes. Represented as a matrix
$I_{ij}$ so that the pairing on two junctions $J_1$
and $J_2$ is $J_{1,i}I_{i,j}J_{2,j}$,
the matrix in this case is given by
\begin{equation}
\begin{aligned}
I &\equiv (\cdot,\cdot)\\
& =\left({\tiny
\begin{matrix}
-1 & 1/2 & 0 & 1/2 & 0 & 1/2 & 0 & 1/2 & 0 & 1/2 & 0 & 1/2 \\
1/2 & -1 & -1/2 & 0 & -1/2 & 0 & -1/2 & 0 & -1/2 & 0 & -1/2 & 0 \\
0 & -1/2 & -1 & 1/2 & 0 & 1/2 & 0 & 1/2 & 0 & 1/2 & 0 & 1/2 \\
1/2 & 0 & 1/2 & -1 & -1/2 & 0 & -1/2 & 0 & -1/2 & 0 & -1/2 & 0 \\
0 & -1/2 & 0 & -1/2 & -1 & 1/2 & 0 & 1/2 & 0 & 1/2 & 0 & 1/2 \\
1/2 & 0 & 1/2 & 0 & 1/2 & -1 & -1/2 & 0 & -1/2 & 0 & -1/2 & 0 \\
0 & -1/2 & 0 & -1/2 & 0 & -1/2 & -1 & 1/2 & 0 & 1/2 & 0 & 1/2 \\
1/2 & 0 & 1/2 & 0 & 1/2 & 0 & 1/2 & -1 & -1/2 & 0 & -1/2 & 0 \\
0 & -1/2 & 0 & -1/2 & 0 & -1/2 & 0 & -1/2 & -1 & 1/2 & 0 & 1/2 \\
1/2 & 0 & 1/2 & 0 & 1/2 & 0 & 1/2 & 0 & 1/2 & -1 & -1/2 & 0 \\
0 & -1/2 & 0 & -1/2 & 0 & -1/2 & 0 & -1/2 & 0 & -1/2 & -1 & 1/2 \\
1/2 & 0 & 1/2 & 0 & 1/2 & 0 & 1/2 & 0 & 1/2 & 0 & 1/2 & -1
\end{matrix}}\right) \,.
\end{aligned}
\end{equation}
With this information, we may compute 
relative homology elements $J$, their boundary,
and the value of two such elements under the
pairing.

Since $H_2(dP_9,\mathbb{Z})$ has a $-E_8$ sublattice,
we wish to recover the $-E_8$ directly. Given the
ordering the we have chosen, where the $10$ $I_1$ loci
that arise from deformations of the Kodaira $II^*$ fiber
are the first ten in the ordering, we expect that
the roots of $E_8$ may be realized by 
$J\in H_2(dP_9,E_p)$ that have $J_{11}=J_{12}=0$.
Furthermore, since the $-E_8$ lattice is in
the full homology and not
just the relative homology, and also the roots are $(-2)$-curves,
we expect that the roots arise as $J$ with $a(J)=0$ and
$(J,J)=-2$; this is standard procedure in the
string junction literature. Indeed,
direct computation yields a set of root junctions
\begin{equation}
R:=\{J=\sum_{i=1}^{12}J_i \, \Gamma_i \in H_2(dP_9,E_p) \,\,\, | \,\,\, a(J) = 0, (J,J)=-2, \,\, \text{and}\,\, J_{11}=J_{12}=0\}
\end{equation}
with precisely $240$ elements. 
A set of simple root junctions is 
\begin{align}
\alpha_1&=(0, 0, 0, 1, -1, -1, 0, -1, 1, 1, 0, 0)\nonumber\\
\alpha_2&=(0, 0, 0, 0, 0, 0, 0, 1, 0, -1, 0, 0)\nonumber\\
\alpha_3&=(0, 0, 0, 0, 0, 0, 1, 0, -1, 0, 0, 0)\nonumber\\
\alpha_4&=(0, 0, 0, 0, 1, 0, -1, 0, 0, 0, 0, 0)\nonumber\\
\alpha_5&=(0, 0, 1, 0, -1, 0, 0, 0, 0, 0, 0, 0)\nonumber\\
\alpha_6&=(0, 1, -2, -1, 0, -1, 1, 0, 1, 1, 0, 0)\nonumber\\
\alpha_7&=(1, -1, 1, 1, 0, 1, -1, 0, -1, -1, 0, 0)\nonumber\\
\alpha_8&=(0, 0, 0, 0, 0, 1, -1, -1, 1, 0, 0, 0) \,,
\end{align}
which generate the positive (negative)
elements of $R$ as non-negative (non-positive) linear combinations. They
also reproduce the $-E_8$
Cartan matrix as
\begin{equation}
(\alpha_i,\alpha_j)=
\footnotesize{\begin{pmatrix}
-2 & 1 & 0 & 0 & 0 & 0 & 0 & 0 \\
1 & -2 & 1 & 0 & 0 & 0 & 0 & 0 \\
0 & 1 & -2 & 1 & 0 & 0 & 0 & 1 \\
0 & 0 & 1 & -2 & 1 & 0 & 0 & 0 \\
0 & 0 & 0 & 1 & -2 & 1 & 0 & 0 \\
0 & 0 & 0 & 0 & 1 & -2 & 1 & 0 \\
0 & 0 & 0 & 0 & 0 & 1 & -2 & 0 \\
0 & 0 & 1 & 0 & 0 & 0 & 0 & -2
\end{pmatrix}} \,,
\end{equation}
and also because they generate half of the elements of $R$ (``positive'' root junctions)
as non-negative linear combinations of the $\alpha_i$, as expected
for simple roots.
Letting $SR=(\alpha_i)$ be the $8\times 12$
matrix of simple root junctions,
\begin{equation}
SR\times I
\end{equation}
is an $8\times 12$ matrix that maps a string junction
to its weight (in $\mathbb{Z}^8$) in the Dynkin basis.

Let us now turn to construct the junctions called $\mathfrak{t}_0$ and
$E$ in section \ref{sec:StringJunk}, which are critical for constructing
$\sigma_{\gamma,0}$ from the thimbles $\mathfrak{t}_\gamma$. For the construction to work such that $\mathfrak{t}_\gamma$
has self-intersection $-1$, we require
\begin{equation}
\mathfrak{t}_0^2 = -1,\,\,\, E^2 = 0, \,\,\, \mathfrak{t}_0\cdot E =1\,,
\label{eqn:juncreqs}
\end{equation}
using the pairing on junctions associated with $I$.

$E$ appears as a loop around all twelve defects, which
after performing appropriate Hanany-Witten moves as depicted
in figure  \ref{fig:Ehwmove} is given by the junction
\begin{equation}
E=(2, 1, 1, 2, -1, 1, -2, -1, -1, -2, 1, -1)\,,
\end{equation}
where the entries are the coefficients of the thimbles, i.e.,
$E=\sum_i E_i \Gamma_i$.
Let us describe the figure  and Hanany-Witten moves.
On the left-hand side, we have a set of loops around individual defects that begins
and ends at $p$. Monodromies are computed
counter-clockwise, so that the
monodromy associated with $\points_+$ turns
$(1,1)$ into $(2,1)$, for example.
Successive loops also induce monodromies,
and the cycle obtained by acting with the
monodromy of each loop is displayed on the far
side of the each loop, oriented counter-clockwise. The
right-hand side of the figure displays the junction obtained from the loop
on the left by successively performing
Hanany-Witten moves. The displayed
numbers are the number of prongs obtained
by trading a loop for a prong (i.e., performing the Hanany-Witten move), and they may be determined uniquely
by the vanishing cycle of the seven-brane the loop, the
charge of the incoming and outgoing
cycles, and charge conservation. Since the monodromy associated with a large loop
around all $(p,q)$ defects \footnote{ This is also the the composition of the monodromy of
the loops in the figure.} is trivial, any cycle is fixed upon traversing the entire loop. However,
a loop in the base that is traversed by
the $(1,0)$ or $(0,1)$ cycle are not acted on by the monodromies associated with
the loops around $\points_-$ and $\points_+$, respectively, and therefore the associated junctions
do not end on $\points_+$ or $\points_-$; in \cite{DeWolfe:1998pr} similar loops were called $\delta_1$ and
$\delta_2$, respectively, since the absence of monodromies means that the cycle can
be pulled through the defect without picking up a prong, so that it is equivalent to
a loop around $11$ of the defects.

Let us now turn to $\mathfrak{t}_0$.
The two natural candidates for $\mathfrak{t}_0$
are the junctions that end only on the defects that do not contribute
to $E_8$, which are represented by $(0,0,0,0,0,0,0,0,0,0,1,0)$
and $(0,0,0,0,0,0,0,0,0,0,0,1)$. Requiring $\mathfrak{t}_0\cdot E = (\mathfrak{t}_0,E)= 1$
fixes
\begin{equation}
\mathfrak{t}_0 = (0,0,0,0,0,0,0,0,0,0,1,0),
\end{equation}
and using the explicit form for $I$ one can verify that all requirements
in \eqref{eqn:juncreqs} are satisfied. $\mathfrak{t}_0$ has asymptotic charge $(0,1)^T$. One can also verify that $E$ and
$\mathfrak{t}_0$ are orthogonal to all of the simple roots $\alpha_i$.

We may now construct $\mathfrak{t}_\gamma$. Any $\gamma$ in the $E_8$ lattice
may be written as
\begin{equation}
\gamma = \sum_i a_i \alpha_i,
\end{equation}
and it has $\gamma^2 = -2n$ for some $n$.
We define
\begin{equation}
\mathfrak{t}_\gamma = \gamma + \mathfrak{t}_0 + n E,
\end{equation}
which is a string junction, i.e., and element of $H_2(dP_9,E_p)$,
that has $\mathfrak{t}_\gamma^2=-1$ and asymptotic charge $(1,0)^T$.

\section{Discussion of Instanton Prefactors in F-theory}
\label{sec:universalprefactorevidence}

As we have discussed, the D3-ED3 instanton zero mode sector gives rise to non-universal
prefactors for the instantons studied in \cite{Donagi:1996yf}. 

In this appendix
we provide an in-depth discussion and some calculations, studying
multiple zero mode sectors and their implications for prefactor universality or non-universality. 
Each zero-mode sector may in principle
cause superpotential zeroes or a changed superpotential structure if additional zero
modes arise on subloci in moduli space. This introduces explicit moduli-dependence
into the prefactors. If any zero-mode sector behaves non-universally across an
ensemble of instantons, the associated prefactors are also necessarily non-universal.
In the specific case of \cite{Donagi:1996yf}, universality was argued for from
the existence of an automorphism that swaps sections, one must see how the zero-mode
sector breaks the automorphism. We organize our discussion with respect to various zero-mode sectors in the F-theory description: the ED3-7 and ED3-D3 modes, respectively.

 The zero mode sector we would first like to consider are those arising
 from ED3-7 strings, and we will see that non-trivial necessary conditions for
 universal prefactors are satisfied. In the dual heterotic
 description these arise from vector bundle zero modes that are counted by
 $h^i(\sigma_{\gamma},\mathcal{E}_X|_{\sigma_{\gamma}}\otimes \CO(-1))$, and they
 should correspond to singularities inside associatives in the M-theory picture.
 A necessary condition for a universal prefactor is that whenever an additional ED3-7 Fermi
 zero mode arises for one instanton, it arises for all instantons. Two types of zero modes that may
 arise are from ED3 intersections with non-abelian seven-branes, or with $I_1$ loci
 inside the non-abelian seven-branes. We
 wish to show that in each case, the intersection structure is the same for all instantons
 in the infinite set.
 \begin{itemize}
 	\item \textbf{Modes from intersection with non-abelian brane stacks.} Moving in moduli space
 	such that the structure group
 	of $\mathcal{E}_X$ on the heterotic side  decreases rank gives rise to a non-trivial gauge group.
 	This is dual to the development of non-abelian seven-branes along {$\widehat{dP_9}$} in the
 	threefold base {$B_{DGW}=\widehat{dP_9}\times \mathbb{P}^1$.} If it were the case that some ED3
 	intersected the non-abelian stack but not others, associated zero modes may only
 	arise in the cases where the intersection exists, and a universal prefactor would
 	be unlikely. However, the curve $\sigma_\gamma$ always sits inside the $dP_9$, and
 	therefore these zero modes may in principle arise for every instanton; i.e., every
 	ED3 we consider intersects the non-abelian seven-brane stack, should one develop.
 	\item \textbf{Modes from intersection with $I_1$ loci inside non-abelian brane stacks.} Here the argumentation
 	is similar: if the number of $ED3$-$I_1$ intersections inside the non-abelian stack were different
 	for different instantons, a universal
 	prefactor would be unlikely. Let us compute the number of such intersections.
 	First, allowing for singular limits we write
 	\begin{equation}
 		\Delta = z^N \tilde \Delta \,,
 	\end{equation}
 	where $z$ is the coordinate normal to the $dP_9$ and $\tilde \Delta$ is the $I_1$
 	locus, which is of class 
 	\begin{equation}
 	[\tilde \Delta]=-12K_{dP_9}
 	\end{equation} upon restriction to the
 	$dP_9$, since the normal bundle is trivial for this base. It is independent 
 	of $N$. This intersects that ED3 at the points $\sigma_\gamma\cdot \tilde \Delta$,
 	which inside $dP_9$ is
 	\begin{equation}
 		\sigma_\gamma\cdot \tilde \Delta = 12 \,.
 	\end{equation}
 	This is independent of $\gamma$, and comes from
 	the $\sigma_0\cdot F$ term in $\sigma_\gamma$.
 \end{itemize}
 In summary, if a non-abelian seven-brane exists, all $\sigma_\gamma$ sit inside of
 it regardless of $N$, and furthermore all intersections of the ED3 with $\tilde \Delta$
 inside of the non-abelian seven-brane occur at twelve points, independent of both
 $\gamma$ and $N$. This is evidence that there may be zero mode universality in the
 $ED3$-$7$ sector.

 On the other hand, the F-theory compactification of \cite{Donagi:1996yf} has no
 background four-form flux, and therefore spacetime-filling D3-branes must be 
 introduced to cancel the D3-brane tadpole arising from the structure of the
 $I_1$-locus. When the D3-branes hit the ED3-instantons, additional Ganor
 strings become light and cause a zero in the prefactor. This introduces a
 moduli dependence into the prefactors of each instanton which in general
 lifts some of the D3-brane moduli space. However, given that there are 
 $12$ D3-branes and an infinite number of sections, there is no position
 for the D3-branes in the partially lifted moduli space that preserved the
 automorphism on the Calabi-Yau; i.e., D3-brane positions mark the Calabi-Yau
 and break the automorphism. This breaks prefactor universality, as discussed in
 the main text.


\providecommand{\href}[2]{#2}\begingroup\raggedright\endgroup

\end{document}